\documentstyle[12pt,epsf]{article}
\newcommand{\Tint}[1]{{\hbox{$\sum$}\!\!\!\!\!\!\int}_{\!\!\!\!#1}}
\newcommand{\la}[1]{\label{#1}}
\newcommand{\be}{\begin{equation}}
\newcommand{\ee}{\end{equation}}
\newcommand{\ba}{\begin{eqnarray}}
\newcommand{\ea}{\end{eqnarray}}
\newcommand{\bi}{\begin{itemize}}
\newcommand{\ei}{\end{itemize}}

\newcommand{\nr}[1]{(\ref{#1})}
\newcommand{\tr}{{\rm Tr\,}}
\newcommand{\hc}{{\rm H.c.\ }}
\newcommand{\nn}{\nonumber \\}
\newcommand{\fr}[2]{{\frac{#1}{#2}}}
\newcommand{\msbar}{\overline{\mbox{\rm MS}}}

\newcommand{\bfp}{{\bf p}}
\newcommand{\muT}{\overline\mu_T}
\newcommand{\<}{\left\langle}
\renewcommand{\>}{\right\rangle}

\def\lsi{\raise0.3ex\hbox{$<$\kern-0.75em\raise-1.1ex\hbox{$\sim$}}}
\def\gsi{\raise0.3ex\hbox{$>$\kern-0.75em\raise-1.1ex\hbox{$\sim$}}}
\newcommand{\lsim}{\mathop{\lsi}}
\newcommand{\gsim}{\mathop{\gsi}}
\makeatletter \@addtoreset{equation}{section} \makeatother
\renewcommand{\theequation}{\arabic{section}.\arabic{equation}}
\begin{document}

\begin{titlepage}
\begin{flushright}
CERN-TH/98-116\\
hep-ph/9804237\\
\end{flushright}
\begin{centering}
\vfill

{\bf THERMODYNAMICS OF NON-TOPOLOGICAL SOLITONS}
\vspace{0.8cm}

M. Laine$^{\rm a,b}$\footnote{mikko.laine@cern.ch} and
M. Shaposhnikov$^{\rm a}$\footnote{mshaposh@nxth04.cern.ch} \\

\vspace{0.3cm}
{\em $^{\rm a}$Theory Division, CERN, CH-1211 Geneva 23,
Switzerland\\}
\vspace{0.3cm}
{\em $^{\rm b}$Department of Physics,
P.O.Box 9, 00014 University of Helsinki, Finland\\}

\vspace{0.7cm}
{\bf Abstract}

\end{centering}

\vspace{0.3cm}\noindent
In theories with low energy supersymmetry breaking, the effective
potential for squarks and sleptons has generically nearly flat
directions, $V(\phi)\sim M^4 (\log(\phi/M))^n$. This guarantees the
existence of stable non-topological solitons, Q-balls, that carry
large baryon number, $B \gg (M/m_p)^4$, where $m_p$ is the proton mass.
We study the behaviour of these objects in a high temperature plasma. We
show that in an {\em infinitely} extended system with a finite density
of the baryon charge, the equilibrium state is not homogeneous and
contains Q-balls at {\em any} temperature. In a system with a finite
volume, Q-balls evaporate at a volume dependent temperature. In the
cosmological context, we formulate the conditions under which Q-balls,
produced in the Early Universe, survive till the present time. Finally,
we estimate the baryon to cold dark matter ratio in a cosmological
scenario in which Q-balls are responsible for both the net baryon
number of the Universe and its dark matter. We find out naturally 
the correct orders of magnitude for $M\sim 1\ldots 10$ TeV:
$\eta\sim 10^{-10} (M/{\rm TeV})^{-2} (B/10^{26})^{-1/2}$.
\vfill
\noindent

\vspace{0.8cm}

\noindent
CERN-TH/98-116\\
April 1998

\vfill

\end{titlepage}

\section{Introduction}
\vfill
\noindent
Take a theory containing scalar fields which carry some unbroken global
U(1) charge~$Q$. Assume that all charged particles are massive. Suppose
that the effective potential for the U(1) charged scalar fields has a
flat direction, $V(\phi)\sim M^4$ at large $\phi$, up to possible
logarithmic terms. Then this theory contains {\em absolutely stable}
non-topological solitons, Q-balls, that have a non-zero value of the global
charge $Q$~\cite{fls}. Flatness of the effective potential at large
$\phi$ is essential for this statement. For $Q \gg 1$, the mass of the
soliton grows as $M_Q \sim M Q^{3/4}$ and becomes smaller than the mass
of a collection of free separated particles with the same charge, $m
Q^{1}$, independent of the relationship between $m$ and $M$. For
theories without flat directions, where $V(\phi)$ grows at large $\phi$
as $\phi^n$ with $n \geq 2$, Q-ball solutions constructed from scalar
fields can exist as well \cite{rosen,c}, but their energy scales with
$Q$ as $A Q^1$, so that the question of their stability depends on the
mass of the lightest particle that carries the charge and on the
(computable) coefficient $A$.

A phenomenologically interesting example
of a system with these properties is provided by supersymmetric
extensions of the Standard Model. Here the role of the U(1) charge is
played by the baryon number, and the role of scalar fields by squarks.
The MSSM with unbroken supersymmetry has a lot of gauge invariant flat
directions with $V(\phi)=0$ carrying baryon number \cite{flat}. If
supersymmetry is broken at a low energy scale $M$, as in gauge-mediated
scenarios~\cite{gauge-mediated}, then the potential is lifted at $\phi
> M$ in a way that $V(\phi)$ stays constant \cite{dvali,gmm}. Thus,
this theory contains stable non-topological solitons which carry baryon
number \cite{ks}. The generic MSSM with other types of SUSY breaking
also contains a lot of Q-ball solutions \cite{ak11,ak10}, but 
with $M_Q \sim M Q^{1}$.

Stable Q-balls in supersymmetric theories have a number of interesting
properties. Since they represent the most economic way of packing the
baryon number, they like to absorb the ordinary matter in the form of
protons and neutrons and to convert baryon number from fermionic baryons
to bosonic baryons (squarks) \cite{dks}. Large stable Q-balls can be
naturally produced \cite{ks} in the Early Universe via the decay of an
Affleck-Dine (AD) condensate \cite{ad} and can contribute to the cold
dark matter \cite{ks}. Experimental signatures of relic dark matter
Q-balls are spectacular \cite{kks}.

At zero temperature, Q-balls are the states with minimum {\em energy}
at a fixed value of the charge $Q$. They do not decay if the mass to
charge ratio $M_Q/Q$ for them is smaller than that for a free particle.
At non-zero temperatures all energy levels of the system are populated
with non-zero probabilities and, thus, the charge can leak out from a
Q-ball to the outer space, or, equally, can be absorbed from the
surrounding plasma. Absorption of the charge into a Q-ball decreases
the energy of the system, while releasing the charge from it increases
the entropy. So, the issue of Q-ball stability gets more complicated than
at zero temperature.

The aim of this paper is to consider Q-ball thermodynamics in theories
with flat directions in the scalar potential. We will see that in these
models systems with a finite density of the baryon number $n_B$ do not
have a
standard thermodynamical limit. For large enough volumes the initially
homogeneous distribution of $n_B \neq 0$ is unstable against Q-ball
formation at any temperature. On the contrary, if the volume of the
system is finite and its charge is fixed, Q-balls cease to exist
(evaporate completely) at temperatures $T>T_Q$, where $T_Q$ is volume
dependent.

In the cosmological context, a mere thermodynamical consideration is
not enough as the system has a finite time scale given by the expansion
rate of the Universe, and the (non-equilibrium) rate of Q-ball
evaporation is essential. We make an estimate of this rate and
determine under which conditions Q-balls created in the Early Universe
survive till present. If the only source of baryon asymmetry of the
Universe is the AD mechanism, then the baryon asymmetry and cold dark
matter may share the same origin \cite{ks}. In this scenario, {\em
both} matter and dark matter are baryonic, though dark matter is
constructed from squarks packed inside Q-balls. We show that in this
case, the baryon to cold dark matter ratio appears to be related to
very few parameters such as the charge of an individual Q-ball, 
the Planck scale, the SUSY breaking scale $M$ and the proton mass.

Initially, Q-balls were found and studied in different models
in Refs.~\cite{fls}--\cite{c}. The generation, evolution, evaporation
and implications of Q-balls at the high temperatures of the Early
Universe were discussed in various models
in~\cite{evaporation}--\cite{emc2}. In particular, in~\cite{gk,foga} it
was found that at high temperatures (but still $T\ll M$) Q-balls tend
to evaporate whereas at low temperatures, they are stable and tend to
grow. The influence of SUSY Q-balls on phase transitions was discussed
in \cite{ak9}. In theories where supersymmetry breaking comes from the
supergravity hidden sector, Q-balls are unstable at zero temperature.
Nevertheless, they can be produced in the Early Universe and are
important for baryogenesis \cite{emc1}. Moreover, they may provide a
natural explanation for the ratio between neutralino dark matter and
the baryonic matter in the Universe \cite{emc2}. However, theories with
flat potentials, specific for theories with low energy supersymmetry
breaking have, to our knowledge, never been analyzed in detail for all
temperatures up to $T\gg M$.

This paper is organized as follows. Essentially, it follows two
distinct
lines: a qualitative discussion of the physically interesting case of
the MSSM with order of magnitude estimates, and a more formal and
quantitative analysis, but mostly applied to a specific toy model. In
Sec.~\ref{sec:formu}, we formulate in general terms the main problems
considered. In Sec.~\ref{sec:qballs}, we discuss the solutions to these
problems on a qualitative level. In Sec.~\ref{sec:model} we perform a
more quantitative analysis, but in a simple renormalizable
supersymmetric toy model with flat directions in the effective
potential. A number of claims made in Sec.~\ref{sec:qballs} are
substantiated in Sec.~\ref{sec:model}. Finally, in Sec.~\ref{sec:rate}
we consider the applications of the results to cosmology in the
realistic case, again on an order of magnitude level. In particular, we
consider here the stability of Q-balls at high temperatures, their
evaporation rate, and their survival in the expanding Universe. We also
estimate the baryon to cold dark matter ratio. Some details related to
Sec.~\ref{sec:model} are in the two Appendices.

\section{The formulation of the problem \la{sec:formu}}

Let us consider a system with some conserved global charge $Q$, such as
$B$ or $L$ (baryon number or lepton number; the $B+L$ anomaly 
is not important for this discussion).The symmetry which
corresponds to this charge is assumed to be unbroken. On a formal
level, the main questions we are going to address are as follows:

1. Assuming that the thermodynamical ground state of the system is
homogeneous, is the conserved charge carried by individual particles in
a plasma, or by a scalar condensate (Q-matter), in analogy with
Bose-Einstein condensation?

2. Is such a homogeneous thermodynamical ground state stable against
small inhomogeneous perturbations? If it is, is it the true ground
state or a only metastable minimum?

3. In case a homogeneous ground state is unstable or metastable, what
is the global minimum of the free energy? In particular, under which
conditions can Q-balls constitute such a ground state?

4. If one starts from an unstable or metastable solution, at which rate
does one approach the equilibrium ground state? In particular, at which
rate do Q-balls form if they are stable, or evaporate if they are
unstable?

In order to answer these questions, let us first fix the basic
definitions used. The thermodynamics of a system with a conserved
charge (the case of zero temperature is incorporated as a special
case) is determined by the grand canonical partition function
\be
Z_\mu = e^{-\beta\Omega(T,V,\mu)} = e^{\frac{V}{T} p(T,\mu)} =
\tr e^{-\beta (\hat H - \mu \hat Q)}, \la{Zmu}
\ee
where
\be
\frac{\<\hat Q\>}{V} \equiv \frac{Q}{V}
\equiv q =
\frac{\partial p(T,\mu)}{\partial\mu}. \la{defofq}
\ee
If $Q$ is kept fixed instead of $\mu$, then the thermodynamical
potential to be minimized is obtained with a Legendre transformation:
\be
F(T,V,Q)=V f(T,q)=\Omega+\mu Q=V[-p+\mu q]. \la{FTVQ}
\ee
The relation inverse to eq.~\nr{defofq} is then
\be
\mu =\frac{\partial f(T,q)}{\partial q}.
\ee

Given the Lagrangian of the system, one can by standard
methods~\cite{kapusta} derive a Euclidian path integral expression for
$Z_\mu$. In a field theory containing the fields $\Phi$, it is
convenient to express the result in terms of an effective action in the
usual way. It then follows that
\be
p(T,\mu) = -(\beta V)^{-1} \left.\Gamma_\mu[\Phi]
\right|_{\delta\Gamma_\mu[\Phi]/\delta\Phi=0}. \la{effaction}
\ee
The relevant extremum is the minimum, as follows from the general
thermodynamical principle that, for fixed $T,\mu$, the other variables
take such values that the corresponding free energy is minimized.

In the case of a homogeneous extremum, the problem reduces to an
effective potential. Then
\be
p(T,\mu) = -\left. V_\mu(\Phi)\right|_{dV_\mu/d\Phi=0}. \la{defVmu}
\ee
The Legendre transform is
$V_q(\Phi)=V_\mu(\Phi)+\mu q$
where $q=-\partial_\mu V_\mu(\Phi)$, and
\be
f(T,q) = \left. V_q(\Phi)\right|_{dV_q/d\Phi=0}. \la{defVq}
\ee

With these tools, the first three of the questions formulated above
can, in principle, be answered. The first of the questions is
particularly straightforward: one just computes the effective potential
for a fixed charge, $V_q(\Phi)$, and inspects whether a non-zero value
for $\Phi$ serves to minimize it or not. If yes, then one can have a
Q-matter condensate.

To address the second question, one can study the standard
thermodynamical stability conditions: these follow from the requirement
that the entropy of any subsystem must be at a maximum with respect to
local fluctuations of temperature, volume, and the number of charged
particles. Applied to the present case, it follows that to be stable a
medium must satisfy
\be
c_V= - T \left(\frac{\partial^2 f}{\partial T^2}\right)_{q} > 0; \quad
\kappa_T = \frac{1}{q} \left(\frac{\partial p}{\partial q}
\right)_T^{-1} > 0; \quad
\left(\frac{\partial \mu}{\partial q}\right)_T =
\left(\frac{\partial^2 f}{\partial q^2}\right)_T > 0. \la{stab1}
\ee
The last two conditions are, in fact, equivalent. The condition for
$c_V$ is trivially satisfied in a relativistic high temperature system,
since the free energy density $f$ is dominated by the term $\sim -T^4$.
Hence essentially only the last condition remains to be inspected.

Let us note that there is an equivalent formulation for the second
question. Indeed, the fluctuation matrix around the saddle point used
in the evaluation of eq.~\nr{effaction} must be positive definite, from
which it follows that
\be
\partial_{\Phi}^2 \left.V_\mu(\Phi)
\right|_{\Phi=\Phi_{\rm min}} > 0. \la{stab2}
\ee

To address the third question, one has to study non-homogeneous
configurations in eq.~\nr{effaction}. To do this in full generality is,
obviously, very difficult. However, when one restricts to spherically
symmetric configurations (Q-balls), which is expected to be the
relevant case, the problem becomes, to a good approximation, solvable.

Finally, the fourth question is a non-equilibrium consideration, and
the tools above do not directly apply. However, the equilibrium limit
turns out to give quite strong constraints. As the results depend on
the initial non-equilibrium state, we will not attempt a general
analysis here, but rather concentrate on the cosmologically interesting
case of Q-ball evaporation in the Early Universe. This will be done in
Sec.~\ref{sec:rate}.

\section{Q-balls at zero and finite temperatures \la{sec:qballs}}

It turns out that the first and second questions formulated in
Sec.~\ref{sec:formu}, concerning the existence and stability of
Q-matter, have a simple and transparent answer: a Q-matter
state ($\sim$ Bose-Einstein condensate) could in principle exist, but
it is unstable and will decay into an inhomogeneous configuration. We
will illustrate this phenomenon in some detail in Sec.~\ref{sec:model}.
In this Section we consider the inhomogeneous final state, i.e.,
Q-balls, in general qualitative terms. Some more details in a specific
model will again be given in Sec.~\ref{sec:model}.

Let us start by reviewing the properties of Q-balls at zero
temperature. Consider a generic field theory containing scalar fields
$\phi_i$ and fermions $\psi_i$ with global U(1) charges~$q_i$. The
scalar potential is $V(\phi)$. Complications appearing with the
inclusion
of gauge fields were discussed in \cite{kst} and do not appear in the
phenomenologically interesting case of SUSY theories, where flat
directions are associated with SU(3)$\times$SU(2)$\times$U(1) singlets.
Then, a spherically symmetric Q-ball solution has the
form~\cite{fls}--\cite{c}
\be
\phi_i(r,t)=\exp(i q_i \omega t) \phi_i(r).
\ee
The functions $\phi_i(r)$ can be found by minimizing the functional
\be
E_\omega = \int d^3x \, \Bigl[\sum_j|\partial_i \phi_j(r)|^2 +
\hat{V}_\omega(\phi)\Bigr],
\label{energy}
\ee
where
\be
\hat{V}_\omega(\phi)=V(\phi) - \omega^2\sum_i q_i^2 \phi_i^\dagger
\phi_i.
\label{effpot}
\ee
The frequency $\omega$ is related to the total charge of the solution
$Q$  as
\be
Q = 2\omega\int d^3x \sum_i q_i^2  \phi_i^\dagger \phi_i.
\label{omega}
\ee

Consider now the finite temperature case and a large but finite volume
V. The thermodynamics of the system is determined by the grand
canonical partition function defined in eq.~(\ref{Zmu}).

Given the Lagrangian of the system, which contains all the fields of
the theory, one can derive a Euclidian path integral expression for
$Z_\mu$,
\be
Z_\mu=\int {\cal D} [{\rm All~fields}]
\exp\left(
-\int_0^\beta\!d\tau\int\! d^3x {\cal L}_E
\right),
\la{intZmu}
\ee
where ${\cal L}_E$ is the ordinary Euclidean action with the
replacement of the Euclidian time derivative ~\cite{kapusta}
\be
\partial_\tau \rightarrow \partial_\tau - q_i \mu.
\ee
Note that for scalar fields, the charges come with the opposite signs
in $\partial_\tau \phi, \partial_\tau \phi^\dagger$. As usual, bosonic
fields are periodic, while fermionic fields are anti-periodic on the
finite time interval $0< \tau < \beta=1/T$.

To define if there are any non-trivial contributions to the partition
function, potentially associated with Q-balls, one can look for the
saddle points of the exponential in eq.~(\ref{intZmu}). The
configurations that play the most important role at high temperatures
are static, i.e., independent of Euclidian time. The static bosonic
part of the action has the form
\be
S = \beta \int d^3x \,\Bigl[\sum_j|\partial_i \phi_j(r)|^2 +
V(\phi) - \mu^2\sum_i q_i^2 \phi_i^\dagger \phi_i\Bigr],
\label{Fq}
\ee
precisely the one given by eqs.~\nr{energy}, \nr{effpot} with the
replacement $\omega \rightarrow \mu$. Thus, if the theory at zero
temperature has Q-ball solutions with some values of $\omega$, these
solutions are saddle points of the Euclidian finite temperature and
finite density path integral at $\mu = \omega$. The relationship
between the charge and $\mu$ is now given by eq.~(\ref{defofq}), rather
than eq.~(\ref{omega}): besides Q-balls, at non-zero temperatures
charge can be carried by particles outside it, as well.

As is very well known from the study of phase transitions at high
temperatures, the saddle points of the {\em tree-level} static bosonic
action provide a good approximation only when $T \ll M$, where $M$ is
the typical mass scale of the theory. At $T \gsim M$, the quantum
corrections are large, and must be resummed in some way. Thus one
either has to use an approximation for the full effective
action $\Gamma[\Phi]$ in eq.~\nr{effaction}, or one has to use
an effective theory where quantum corrections are small. The latter
could be constructed for static but space dependent
bosonic configurations, integrating out all fermions, all
heavy (mass $\gg T$) bosons, and the non-zero
Matsubara frequencies of light bosons.
For static bosonic configurations,
depending on space coordinates only weakly (we will see that this is
indeed the case for sufficiently large Q-balls), the derivative
expansion is most helpful:
\be
S_{\rm eff} = \beta \int d^3x \,\Bigl[\sum_j Z_j(\phi,T,\mu)|\partial_i
\phi_j(r)|^2 +
V_{\rm eff}(\phi,T,\mu)+...\Bigr],
\label{seff}
\ee
where $V_{\rm eff}(\phi,T,\mu)$ is basically the effective potential of
the system at non-zero $T$ and $\mu$ (more precisely, $V_{\rm eff}$ is
the contribution to the effective potential from the degrees of freedom
that have been integrated out). The effective potential as a function
of $\mu$ has been computed in a number of theories and discussed in
connection with the question of the influence of charge density on the
symmetry behaviour in a number of papers \cite{linde}--\cite{bbdw}. If
the
particle masses in the background of the field $\phi$ are smaller than
the
temperature, then the high temperature expansion can be used for the
construction of the effective action precisely along the lines of
Refs.~\cite{generic}--\cite{erice}. If not (as is in fact for
many of the considerations below),
then an effective action incorporating both high and
low temperature behaviour must be used. Now, we have at hands enough
for a qualitative discussion and order of magnitude estimates of the
properties of SUSY Q-balls at high temperatures. For this aim we take
the wave function normalizations in eq.~(\ref{Fq}) to be unity. As a
number of studies \cite{bbfh,generic} show, higher order corrections
in the wave function normalizations are not essential numerically.

Let us take some field $\phi$ representing, in somewhat loose terms,
a gauge-invariant combination of squarks and sleptons
in the MSSM, and carrying baryon (and, perhaps, lepton) number
(for simplicity, we take the charge $q=1$).
Specific examples can be found in \cite{flat}. Under the global baryon
symmetry, the field $\phi$ is transformed as $\phi \rightarrow \exp(i
\alpha) \phi$. Assume that the effective potential is flat along this
direction as shown in Fig. 1,
\be
V(\phi)\rightarrow M^4
\ee
at $\phi\rightarrow \infty$. Then,
at zero temperature the minimum of the energy at a fixed
baryonic charge $Q \gg 1$ corresponds to a time dependent Q-ball
solution which can in the limit of a large charge be written as
\be
\phi(r) \approx \exp(i \omega t)\times\left \{
\begin{array}{ll}
\phi_{Q} (\sin \omega r)/\omega r, \ & r < R_{Q} \\  & \\
0, \ & r \ge R_{Q}.
\end{array}
\right.
\label{ansatz}
\ee
Here
\be
R_{Q}  \approx  (1/\sqrt{2})\,  M^{-1} \,  Q^{1/4},~
\phi_{Q}  \approx (1/\sqrt{2}) \,  M \, Q^{1/4},~
\omega R_{Q}\approx\pi \la{props1},
\ee
and the mass of the solution is
\be
M_{Q}  \approx (4\pi \sqrt{2}/3) \, M \, Q^{3/4}. \la{props2}
\ee
A Q-ball is stable (cannot decay into protons: $M_Q < m_p Q$)
if $Q>10^{15} (M/1 {\rm TeV})^4$.

Let us now see what happens at high temperatures. The scalar field
$\phi$ couples to the other fields of the MSSM. In the
background of this field a number of particles (e.g. gluons and
gluinos) acquire masses $m(\phi) \sim g \phi$, while others do not
(e.g. if the Q-ball solution at $T=0$ is constructed from squarks only,
leptons and sleptons remain massless since they do not have tree level
interactions with squarks). Now, if $g \phi > T$, massive particles do
not contribute to the finite temperature effective potential just
because of the Boltzmann suppression $\exp(-m(\phi)/T)$, whereas the
contribution of the massless particles is $\phi$ independent and
produces just an overall shift of the effective potential which we
discard. At the same time, $m(\phi)$ is small near $\phi=0$. Thus, the
point $\phi=0$ moves down as $\Delta V \sim -N T^4$, where the
coefficient $N$ is related to the difference between the number of
light degrees of freedom at small and large $\phi$. So, the finite
temperature effective potential has qualitatively the same form as the
effective potential at zero temperature, with the change $M^4
\rightarrow M(T)^4 \sim M^4 + N T^4$.

This consideration allows to write down immediately the Q-ball
contribution to the thermodynamic potential. Replacing $M_Q\to F_Q$,
$M\to M(T)$ in eq.~\nr{props2} and using that $Q^{1/4}\sim R_Q M\sim
M/\omega\sim M/\mu$, we get
\be
F_Q(\mu)\sim M(T) \left(\frac{M(T)}{\mu}\right)^3,
\ee
with the charge in the Q-ball
\be
Q_Q(\mu) \sim \left(\frac{M(T)}{\mu}\right)^4.
\ee
The size of the Q-ball is given by $R_Q\sim \pi/\mu$.

Now, the $\mu$-dependent part of the thermodynamic potential has the
form
\be
\Omega(T,V,\mu) = F_Q(\mu) - C\frac{1}{2} \mu^2 T^2 V,
\label{Omega}
\ee
where the coefficient $C$ is related to the number of light carriers of
the baryon charge (such as quarks). The total charge of the system
follows from eq.~(\ref{defofq}),
\be
Q_{\rm total}=  Q_Q(\mu) + C \mu T^2 V.
\label{Q}
\ee
These two equations define equilibrium Q-ball properties at high
temperatures.

Let us assume first that the average charge density $n_Q$ is nonzero
and is fixed, while the volume of the system grows. Then eq.~(\ref{Q})
admits two solutions. The first one is with a ``large" chemical
potential $\mu \sim n_Q/(C T^2)$. It corresponds to a state where all
the charge is carried by particles in the plasma and Q-balls are not
present. Another solution has a ``small" chemical potential,
\be
\mu \sim M(T)(n_Q V)^{-1/4}.
\ee
It corresponds to a  system containing one Q-ball that carries almost
all of the charge $Q_{\rm total}= n_Q V$.

Constructing the Legendre transform of eq.~\nr{Omega} according to
eq.~\nr{FTVQ}, it can now be seen that the latter solution has a
smaller free energy at large volumes. Indeed, the Q-ball free
energy scales with volume only as $F_Q \sim V^{3/4}$, whereas the free
energy of the plasma phase grows as $F_{\rm plasma}\sim V$. Thus, we
arrive at the conclusion that in theories with flat potentials and a
fixed non-zero density of the charge $Q$, the ground state of the
system always contains just one Q-ball at any temperature when $V
\rightarrow \infty$.

Let us then take another limit, where the total charge of the system is
fixed, and we increase its volume. In this case, the average density of
the charge decreases as $n_Q \sim Q_{\rm total}/V$. Then,
a state with a Q-ball is more favourable than a
homogeneous distribution of the charge at
\be
V \lsim \frac{Q_{\rm total}^{5/4}}{M(T)T^2}.
\label{volume}
\ee
This inequality allows, for example, an estimate of the temperature
$T_{\rm evap}$ at which an initially cold Q-ball, placed in a volume
$V$, evaporates. It follows from eq.~(\ref{volume}) that $T_{\rm evap}
\sim V^{-1/3} Q^{5/12}$, provided $T_{\rm evap} \gg M$, and $T_{\rm
evap} \sim (VM)^{-1/2} Q^{5/8}$ in the opposite case.

All these results have a simple physical meaning. Imagine that we place
a Q-ball at zero temperature into a box with the volume $V$. Now, let
us
heat the system gradually without adding any charge to it. Some charge
from the Q-ball ($\delta Q$) will evaporate, and it will create a
chemical potential in the surrounding plasma, $\mu_{\rm plasma} \sim
\delta Q/(V T^2)$. The process of Q-ball evaporation will stop when the
chemical potential associated with the Q-ball, $ \mu_Q \sim M(T)
(Q-\delta Q)^{-1/4}$, will be equal to $\mu_{\rm plasma}$. If the
inequality in eq.~(\ref{volume}) is satisfied, then one finds that
the solution for $\delta Q$ is
\be
\frac{\delta Q}{Q} \sim \frac{M(T) V T^2}{Q^{5/4}} \ll 1,
\ee
so that the Q-ball cannot evaporate completely, whereas if
eq.~(\ref{volume}) is not true, then the equation $\mu_{\rm
plasma}=\mu_Q$
does not have physical solutions with $Q-\delta Q\gg 1$, and therefore,
the Q-ball disappears.

As we see, the structure of the ground state is obviously volume
dependent. Thus, the system we consider does not have a standard
thermodynamic limit. The flatness of the potential is essential for
these conclusions. If $F_Q \sim Q^1$ rather than $F_Q \sim Q^{3/4}$,
then the standard thermodynamical limit is well defined, and Q-balls
cease to exist above some temperature which is entirely determined by
the density of the charge and the parameters of the model.

\section{A model computation \la{sec:model}}

In order to illustrate in more specific terms the issues discussed
above, we will in this Section consider in some detail the
renormalizable model introduced in~\cite{dvali} for studying
supersymmetric inflation. We first discuss the stability of Q-matter,
and then that of Q-balls, adding numerical coefficients to the
estimates in Sec.~\ref{sec:qballs}.

The model is defined by a superpotential with two complex fields
$\phi,\chi$:
\be
W=\fr12 f \chi\phi^2-\chi m^2.
\ee
The parameters $f,m^2$ can be assumed to be real. We denote $\tilde
m^2= fm^2$. In the direction of the charged field, the potential of the
model grows only logarithmically. Let us assume that there are no local
symmetries and thus no gauge fields.

We start by writing down the Minkowskian Lagrangian. Let us split
$\phi$ to real components, $\phi=(\phi_1+i \phi_2)/\sqrt{2}$, and
denote the Weyl spinors corresponding to $\phi,\chi$ by $\psi_\phi$,
$\psi_\chi$. Then
\ba
{\cal L}_M & =  &
\partial^\mu\phi^*\partial_\mu \phi+
\partial^\mu\chi^*\partial_\mu \chi+
i\overline\psi_\phi\overline\sigma^\mu\partial_\mu\psi_\phi+
i\overline\psi_\chi\overline\sigma^\mu\partial_\mu\psi_\chi-
V(\phi_1,\phi_2,\chi)-{\cal L}_\psi, \hspace*{0.5cm} \la{Mac}
\ea
where
\ba
V(\phi_1,\phi_2,\chi) & =  &\fr12 \tilde m^2(\phi_2^2-\phi_1^2)+
\frac{1}{16} f^2 (\phi_1^2+\phi_2^2)^2+
\fr12 f^2(\phi_1^2+\phi_2^2)|\chi|^2+\frac{\tilde m^4}{f^2},
\hspace*{0.5cm} \la{MV} \\
{\cal L}_\psi & = & \fr12 f\chi \psi_\phi\psi_\phi +
f\phi\psi_\chi\psi_\phi + \hc \la{MLpsi}
\ea
This action has a global U(1)-symmetry,
\ba
 & & \phi\to\phi,\quad\quad \chi\to e^{i\alpha} \chi, \nonumber \\
 & & \psi_\phi\to e^{-i\alpha/2}\psi_\phi, \quad\quad
\psi_\chi \to e^{i\alpha/2}\psi_\chi.
\ea
The superpotential is not symmetric and therefore the scalar fields and
the corresponding fermions transform differently. The charge
corresponding to the symmetry is
\be
Q=\int\! d^3x\left[
i(\chi^*\partial^0\chi-\chi\partial^0\chi^*)+
\fr12\overline\psi_\chi\overline\sigma^0\psi_\chi-
\fr12\overline\psi_\phi\overline\sigma^0\psi_\phi\right]. \la{charge}
\ee
The theory has also a discrete symmetry,
$\phi\to-\phi,\psi_\phi\to-\psi_\phi$, which however does not play any
role in the following.

At tree-level at zero temperature, the ground state of the system is at
a broken minimum: $\chi=0,\phi_2=0,\phi_1^2=4\tilde m^2/f^2,
V(\phi,\chi)=0$. There supersymmetry is conserved, and the spectrum
consists of four massive scalar degrees of freedom and one Dirac
fermion, all with the same mass, $\sqrt{2}\tilde m$. On the other hand,
for large values of $|\chi|$, the minimum of the potential is at
$\phi_1=\phi_2=0$ and the tree-level potential does not depend on
$\chi$: this is a flat direction. Along the flat direction, two scalar
degrees of freedom and one Majorana fermion remain massless. The value
of the potential along the flat direction is $\tilde m^4/f^2= m^4$.

Consider then $Z_\mu$ in eq.~\nr{Zmu}. According to the standard
procedure~\cite{kapusta}, the Euclidian path integral expression for
$Z_\mu$ with the charge $Q$ as given in eq.~\nr{charge}, is
\be
Z_\mu=\int_{\rm b.c.}
{\cal D}\phi
{\cal D}\chi
{\cal D}\psi_\phi
{\cal D}\psi_\chi
\exp\left(
-\int_0^\beta\!d\tau\int\! d^3x {\cal L}_E
\right), \la{iZmu}
\ee
where, denoting now also $\chi=(\chi_1+i\chi_2)/\sqrt{2}$,
the Euclidian Lagrangian is
\ba
{\cal L}_E
& = &
\fr12(\partial_\tau\phi_1)^2+
\fr12(\partial_\tau\phi_2)^2+
\fr12(\nabla\phi_1)^2+
\fr12(\nabla\phi_2)^2 \nonumber \\
& + &
\fr12(\partial_\tau\chi_1-i\mu\chi_2)^2+
\fr12(\partial_\tau\chi_2+i\mu\chi_1)^2+
\fr12(\nabla\chi_1)^2+
\fr12(\nabla\chi_2)^2 \nonumber \\
& + &
\overline\psi_\phi\overline\sigma^0\left(\partial_\tau+\frac{\mu}{2}
\right)\psi_\phi-
i\overline\psi_\phi\overline\sigma^i\partial_i \psi_\phi \nonumber \\
& + &
\overline\psi_\chi\overline\sigma^0\left(\partial_\tau-\frac{\mu}{2}
\right)\psi_\chi-
i\overline\psi_\chi\overline\sigma^i\partial_i \psi_\chi  \nonumber \\
& + & V(\phi,\chi)+{\cal L}_\chi. \la{EuclL}
\ea
Here $V(\phi,\chi)$, ${\cal L}_\chi$ are from eqs.~\nr{MV}, \nr{MLpsi}.

We now wish to consider the functional integral for $Z_\mu$ in the
background of $\hat\phi_1,\hat\chi_1$, i.e., to compute the effective
potentials $V_\mu(\hat\phi_1,\hat\chi_1)$,
$V_q(\hat\phi_1,\hat\chi_1)$,
defined in eqs.~\nr{defVmu}, \nr{defVq}.
For $\chi$, choosing to consider $\chi_1$
is no restriction, due to the U(1) symmetry. On the other hand,
we have $\phi_2=0$ already at the tree-level at zero temperature, so
$\phi_2$ remains at origin for $T>0$. In the following, we denote
$\hat\phi_1=\phi_1,\hat\chi_1=\chi_1$.

The details of the computation of the 1-loop effective potential are
discussed in Appendix A. We consider here the numerical results and
their main features. To have a weakly coupled theory, we take
$f = 0.5$.

\subsection{The stability of Q-matter}

%%%%%%%%%%%%%%%%%%%%%%%%%%%%%%%%% FIGURE
\begin{figure}[t]

\vspace*{-1cm}

\centerline{ %\hspace{-3.3mm}
\epsfxsize=9cm\epsfbox{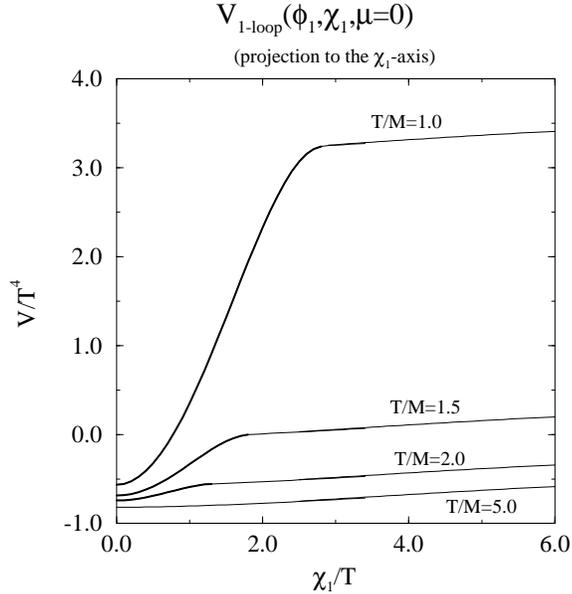}}

\vspace*{-4.5cm}

\caption[a]{The projection of the effective potential on the
$\chi_1$-axis, at different temperatures and $\mu=0$. The temperature
is given in units of the mass scale $M\equiv\tilde m_0 =\tilde m(\tilde
m_0)$. The thick lines indicate the regime where $\<\phi_1\>\neq 0$:
this symmetry is restored at $T/\tilde m_0\gsim 4.0$. For $\chi_1/T\lsim
3.5$, the effective potential is the one of the 3d effective theory in
eqs.~\nr{3dtree}, \nr{3d1loop}, and for $\chi_1/T\gsim 2.5$, the one of
the 4d theory in eqs.~\nr{4dtree}, \nr{4d1loop}. The scale of $\chi_1$
is in both cases what appears in the 3d theory: $\chi_1\equiv
\chi_1^{\rm 4d}(\overline\mu_T/4)$.}
\la{fig:1}
\end{figure}
%%%%%%%%%%%%%%%%%%%%%%%%%%%%%%%%%%%%

%%%%%%%%%%%%%%%%%%%%%%%%%%%%%%%%% FIGURE
\begin{figure}[t]

\centerline{\hspace{-3.3mm}
\epsfxsize=9cm\epsfbox{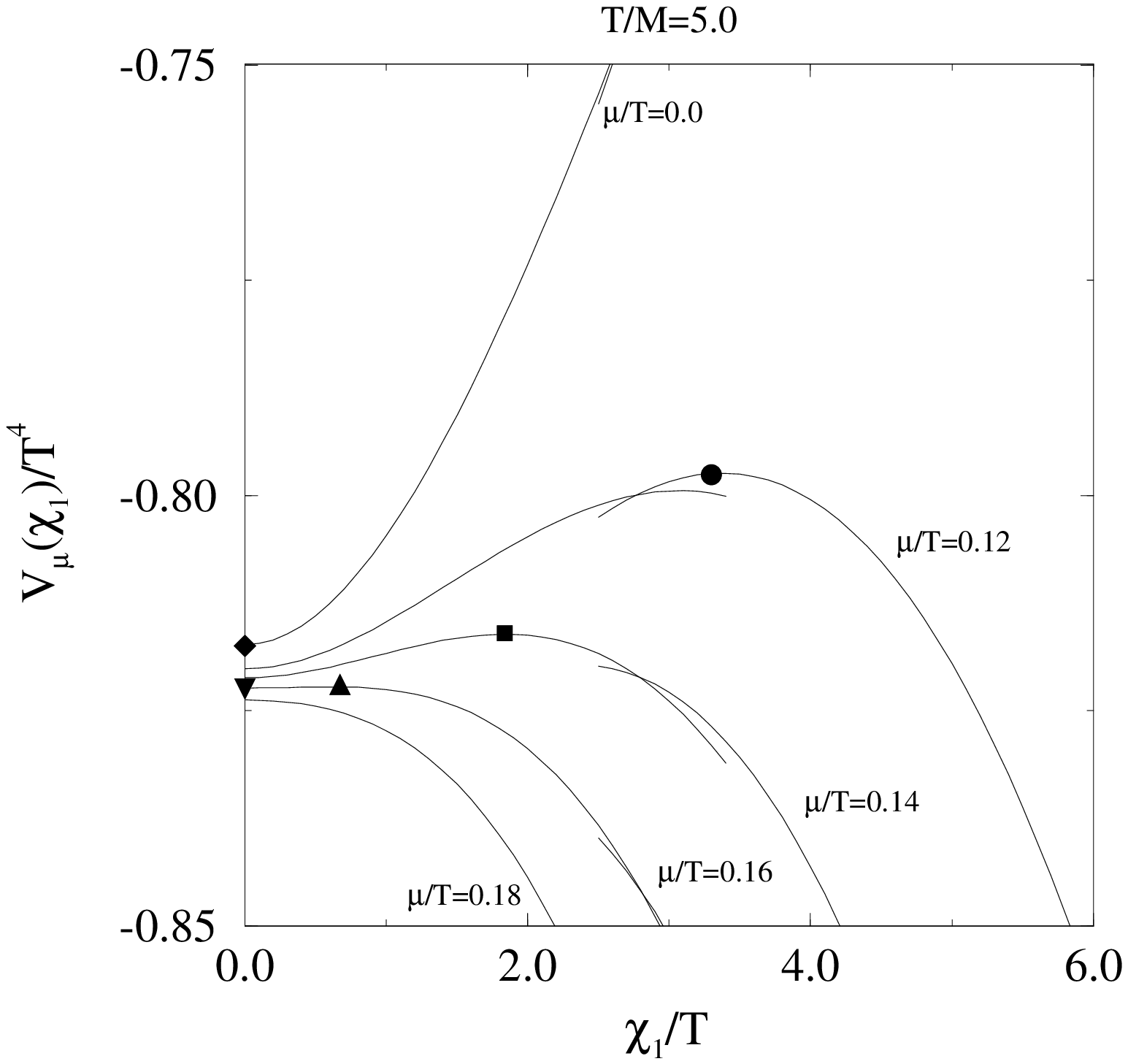}
\hspace{-1.5cm}
\epsfxsize=9cm\epsfbox{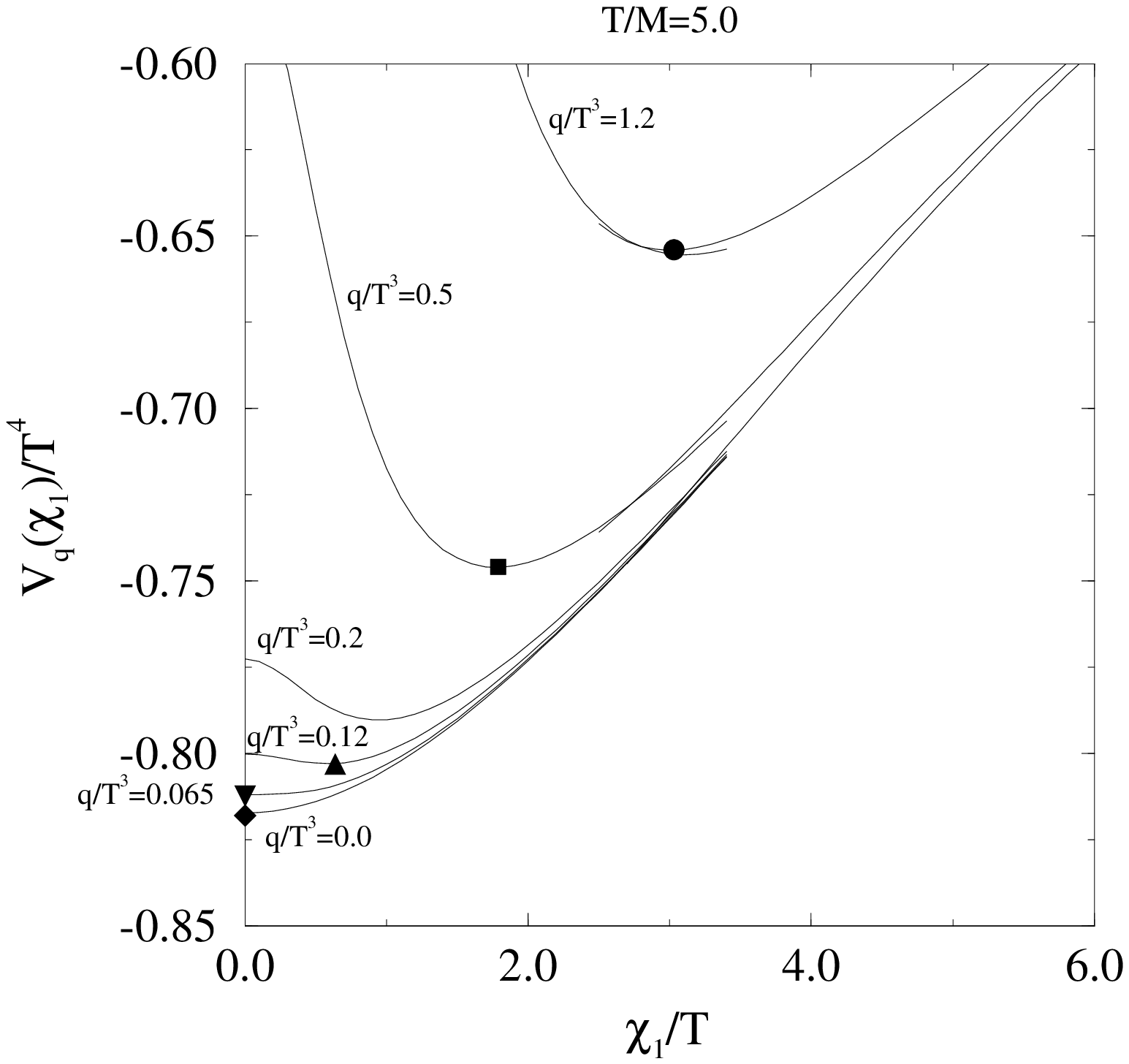}}

\vspace*{-4.5cm}

\caption[a]{
{\em Left:}
The effective potential $V_\mu(\chi_1)$ at $T/\tilde m_0=5.0$, for
different values of $\mu/T$. {\em Right:} The effective potential
$V_q(\chi_1)$ at $T/\tilde m_0=5.0$, for different values of $q/T^3$.
Each extremum of $V_\mu(\chi_1)$ corresponds to an extremum of
$V_q(\chi_1)$, as well, as is shown by the symbols. For
$q/T^3>(q/T^3)_c\sim 0.065$, there is a charged condensate:
$\<\chi_1\>>0$.}
\la{fig:2}
\end{figure}
%%%%%%%%%%%%%%%%%%%%%%%%%%%%%%%%%%%%

%%%%%%%%%%%%%%%%%%%%%%%%%%%%%%%%% FIGURE
\begin{figure}[t]

\centerline{\hspace{-3.3mm}
\epsfxsize=9cm\epsfbox{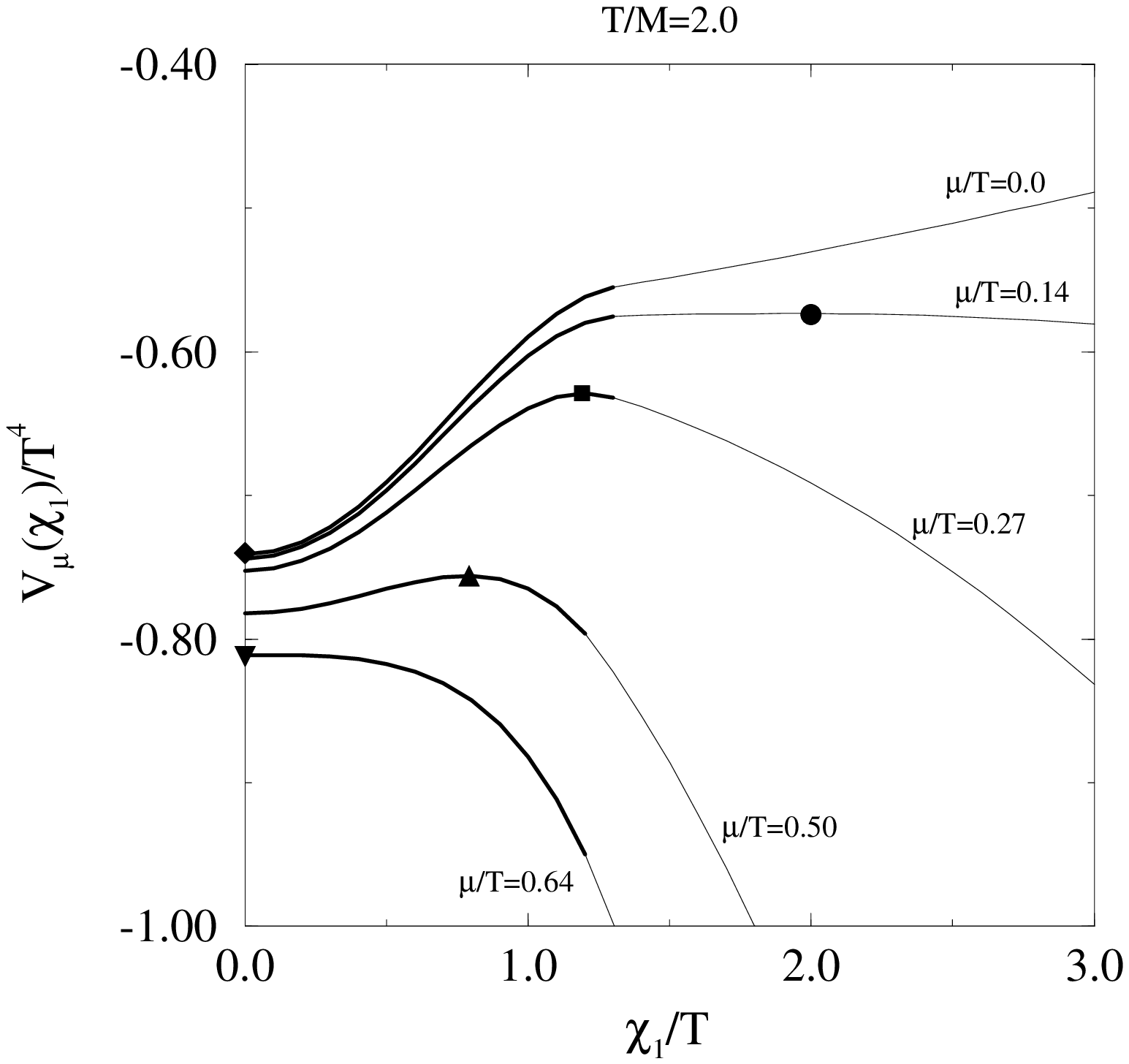}
\hspace{-1.5cm}
\epsfxsize=9cm\epsfbox{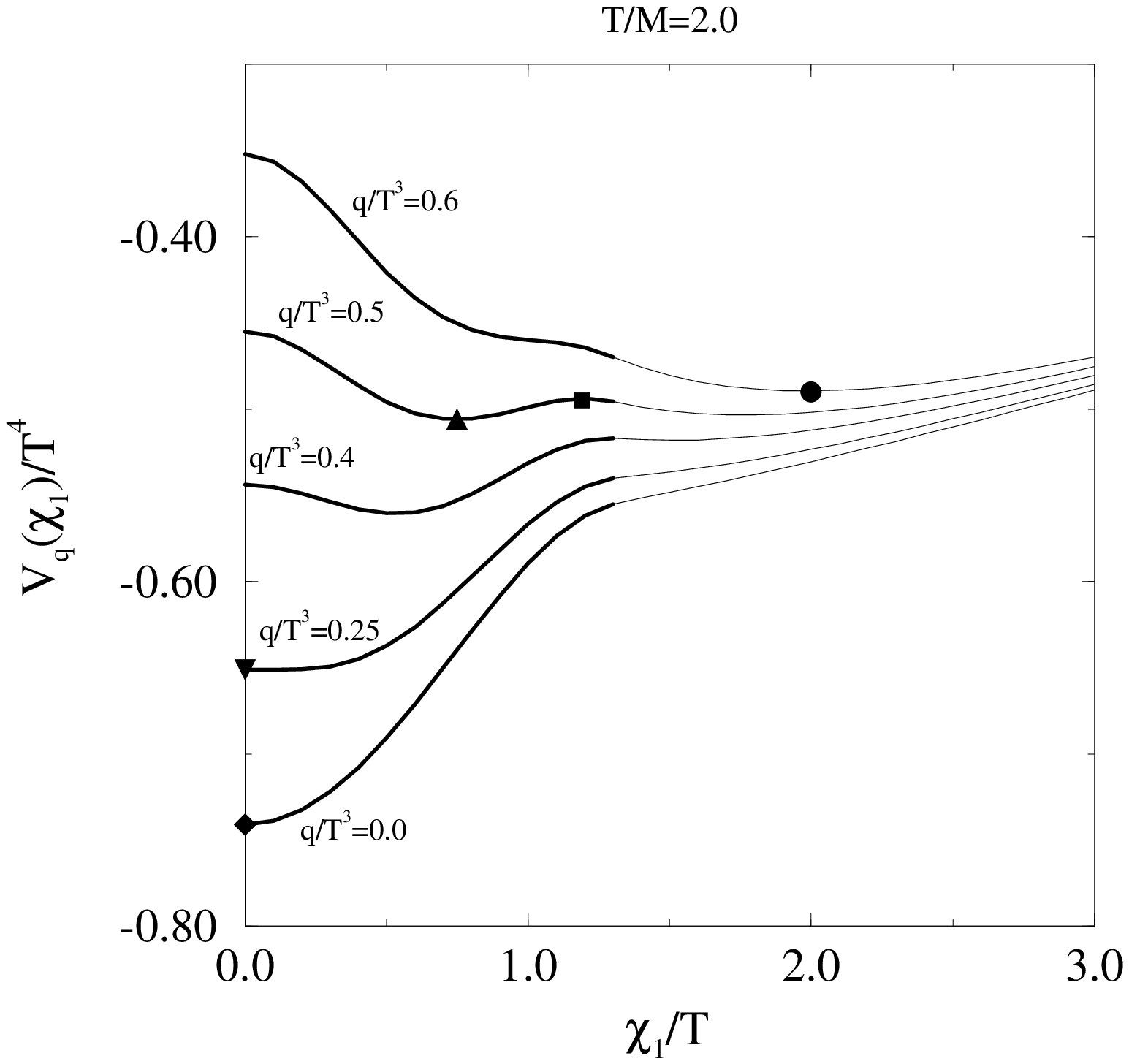}}

\vspace*{-4.5cm}

\caption[a]{
As Fig.~\ref{fig:2}, but for $T/\tilde m_0=2.0$. The thick lines
indicate the regime where the symmetry is broken, $\<\phi_1\>\neq 0$.
Due to the symmetry breaking, the potentials have more structure than
in Fig.~\ref{fig:2}. The critical charge density is here
$(q/T^3)_c\approx 0.25$. Note that the (unique) extremum of
$V_\mu(\chi_1)$ at $\chi_1>0$ is always a maximum. }
\la{fig:3}
\end{figure}
%%%%%%%%%%%%%%%%%%%%%%%%%%%%%%%%%%%%

%%%%%%%%%%%%%%%%%%%%%%%%%%%%%%%%% FIGURE
\begin{figure}[t]

\vspace*{-1cm}

\centerline{ %\hspace{-3.3mm}
\epsfxsize=9cm\epsfbox{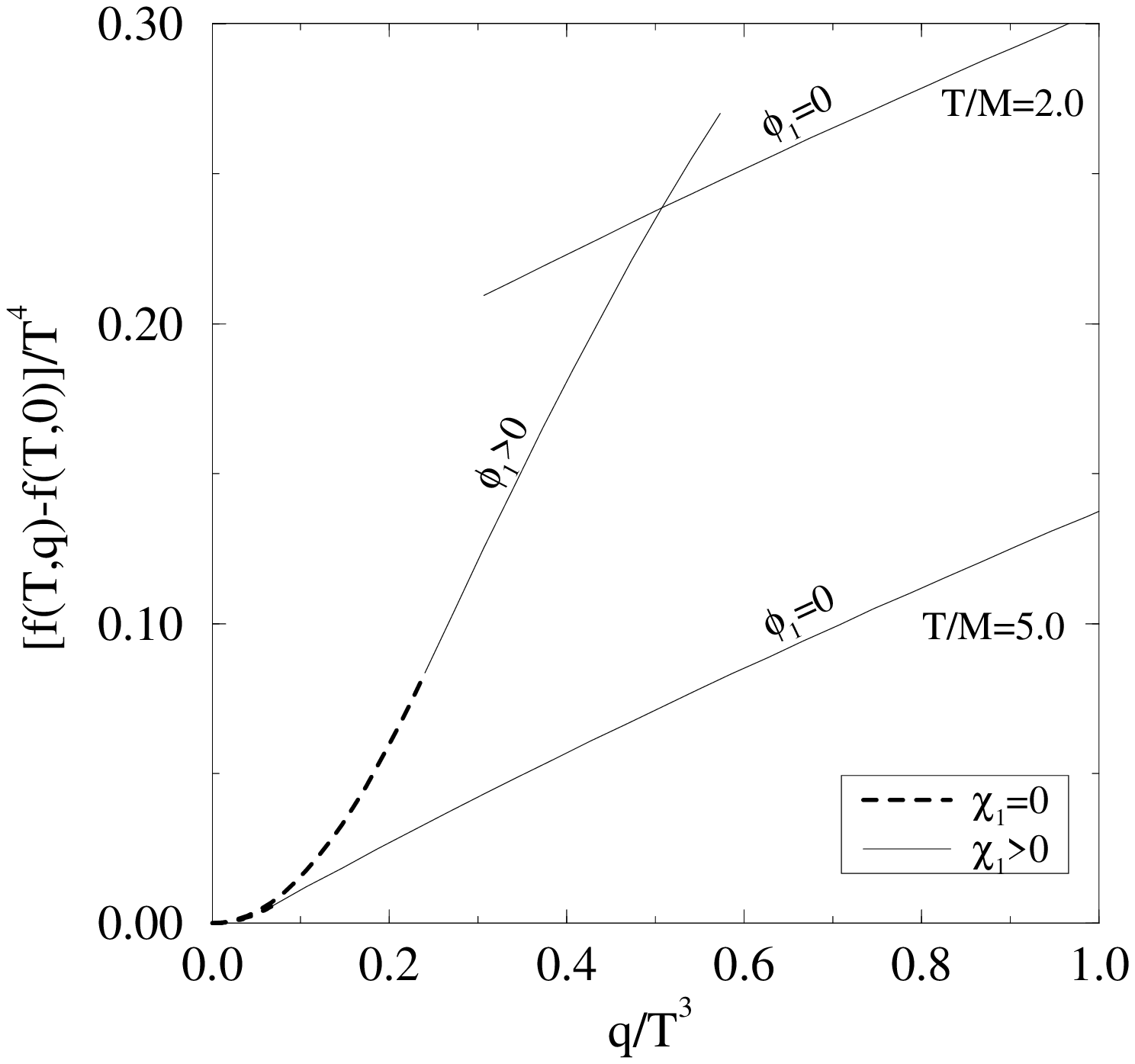}}

\vspace*{-4.5cm}

\caption[a]{The free energy density as a function of the charge density
for $T/\tilde m_0=2.0,5.0$. For $T/\tilde m_0\gsim 4$ the symmetry is
always restored ($\phi_1=0$) and the results scale with $T$; the
critical charge density below which $\chi_1=0$, is $(q/T^3)_c\approx
0.065$. For temperatures $T/\tilde m_0 \lsim 4$, on the other hand, the
symmetry is broken ($\phi_1>0$) at small $\chi_1$, and there is more
structure.}
\la{fig:4}
\end{figure}
%%%%%%%%%%%%%%%%%%%%%%%%%%%%%%%%%%%%

In Fig.~\ref{fig:1}, the 1-loop effective potential is shown at
$\mu=0$. It is seen that for $T/\tilde m_0 \gsim 4$, the symmetry is
restored ($\phi_1=0$) for any $\chi_1$. However, irrespective of the
temperature, the effective potential always displays the characteristic
form allowing Q-balls: at small $\chi_1$, there is a thermal or vacuum
mass term, but at large $\chi_1$, the potential flattens off and grows
eventually only logarithmically.

Consider then the case of a finite charge density. In
Figs.~\ref{fig:2}, \ref{fig:3}, the potentials $V_\mu(\chi_1)$ and its
Legendre transform $V_q(\chi_1)$ are shown at $T/\tilde m_0 =5.0, 2.0$,
respectively. The structure is simpler at $T/\tilde m_0=5.0$, where the
$\phi_1$-symmetry is restored everywhere and the mass scale $\tilde
m_0$
does not have much effect. Nevertheless, both cases show the same
general behaviour. There is a critical charge density $q_c>0$. For
$q<q_c$, the minimum of $V_q(\chi_1)$ is at $\chi_1=0$ and the charge
is hence carried by particle excitations in a plasma phase. However,
for $q>q_c$, a phenomenon analogous with Bose-Einstein condensation
takes place, and $\chi_1>0$.

This pattern can be understood already from the tree-level potential of
the effective 3d theory in eq.~\nr{3dtree}. Consider first the
temperatures $T > T_c^{\phi}\approx 2 \tilde m/f \sim 4 \tilde m$. Then
$m_1^2>0$ according to eq.~\nr{3dprms}, and $\phi_1=0$ for all
$\chi_1$. The effective potential $V_\mu(\chi_1)$ is
\be
V_\mu(\chi_1) \stackrel{f\chi_1\ll T}{\approx}
{\rm const.}
 -\frac{5}{24}\mu^2T^2+
\fr12 \left(-\mu^2+\frac{f^2T^2}{8}
\right) \chi_1^2. \la{BEVmu}
\ee
It will be useful to consider a more general
expression for the latter term
in the coefficient of $\chi_1^2$, so let us
replace $f^2T^2/8\to \tilde M^2$. Then the situation looks precisely
like relativistic Bose-Einstein condensation in a free theory.

To proceed, let us make a Legendre transformation into
$V_q(\chi_1)=V_\mu(\chi_1)+\mu q$, where $q=-\partial_\mu
V_\mu(\chi_1)$:
\be
V_q \sim \fr12 \frac{q^2}{(5/12)T^2+\chi_1^2}
+\fr12 \tilde M^2 \chi_1^2. \la{BEVq}
\ee
What then remains is to minimize $V_q(\chi_1)$ with respect to
$\chi_1$, for different $q,T$. The main results are as follows. The
global minimum is at $\chi_1=0$ if $\tilde M>(12/5)(q/T^2)$. Hence, the
critical charge density is
\be
q_c = \frac{5}{12} \tilde M T^2.
\ee
If $q<q_c$, then
\be
\chi_1 = 0;\quad \mu = \frac{12}{5} \frac{q}{T^2}; \quad
f(T,q)=f(T,0)+
\frac{6}{5} \frac{q^2}{T^2}; \quad \partial_{\chi_1}^2 V_\mu(\chi_1)>0.
\la{plprops}
\ee
If $q>q_c$, on the other hand, then
\be
\chi_1 = \sqrt{\frac{q-q_c}{\tilde M}};\quad
\mu = \tilde M; \quad
f(T,q)=f(T,0)+
\fr65\frac{q_c}{T^2}(2 q - q_c); \quad
\partial_{\chi_1}^2 V_\mu(\chi_1)=0. \la{beprops}
\ee
Replacing now $\tilde M^2 \to f^2T^2/8$,
we see that for any given $T > T_c^{\phi}\sim 2\tilde m/f$,
there is a critical charge density
\be
\left(\frac{q}{T^3}\right)_c=\frac{5f}{24\sqrt{2}},
\ee
below which there is no condensate, $\chi_1=0$. Note that for
$T>T_c^{\phi}$, the leading order results in eqs.~\nr{plprops},
\nr{beprops} do not depend on $\tilde m$ at all, and all the
dimensionful quantities scale with $T$. Numerically, $(q/T^3)_c \sim
0.07$, in accordance with Fig.~\ref{fig:2}.

For $T\lsim T_c^{\phi}$, on the other hand, $m_1^2 < 0$ and $\phi_1^2 >
0$ for small $\chi_1$. The potential in eq.~\nr{3dtree} must thus first
be minimized with respect to $\phi_1^2$. After that has been done, the
potential $V_\mu(\chi_1)$ is again of the form in eq.~\nr{BEVmu} for
small $\chi_1$, but now with $\tilde M^2 = 2 \tilde m^2 - 3 f^2T^2/8$.
Thus the mass scale $\tilde m$ appears in the results. For example,
\be
\left(\frac{q}{T^3}\right)_c = \frac{5}{12} \sqrt{\frac{2 \tilde
m^2}{T^2}-
\frac{3 f^2}{8}}, \la{qclowT}
\ee
which gives $(q/T^3)_c\sim 0.27$ for $T/\tilde m= 2.0$,
in agreement with Fig.~\ref{fig:3}.

To summarize, based on free energy considerations alone, one would say
that for any temperature, there is a charge density $q_c$ such that for
$q>q_c$, one has a Bose-Einstein condensate, or a ``Q-matter'' state.
For $T\to 0$, $q_c\to 0$ according to eq.~\nr{qclowT}, and the plasma
phase does not exist at all.

Let us then consider whether such a Q-matter condensate would be
stable. Applying the stability conditions in eqs.~\nr{stab1},
\nr{stab2}, one sees that in the tree level case, the plasma phase
[eq.~\nr{plprops}] is stable while the situation in the condensate
phase [eq.~\nr{beprops}] is marginal: $\partial_q^2
f(T,q)=\partial_{\chi_1}^2 V_\mu(\chi_1)=0$. Hence the system is
stabilized or destabilized through interactions. Adding an interaction
of the type $\lambda (\chi^*\chi)^2$~\cite{bec,bbdw}, in fact
stabilizes the condensate. On the other hand, it is easy to see that in
the present case, the interactions destabilize the condensate!

The fact that the condensate in the present system is unstable, can be
observed from Figs.~\ref{fig:2}, \ref{fig:3}. The extremum
corresponding to the condensate is always a maximum of $V_\mu(\chi_1)$,
and thus the inequality in eq.~\nr{stab2} is violated. Equivalently,
one can observe from Fig.~\ref{fig:4} that the free energy density
$f(T,q)$ has a negative curvature at $q>q_c$, which violates the
stability condition in eq.~\nr{stab1}. The negative curvature can be
understood also analytically: at large charges, corresponding to large
$\chi_1$'s,
\ba
V_\mu(\chi_1) & \stackrel{f \chi_1\gg T}{\approx} &
V_\mu(0) + \frac{\tilde m^4}{f^2}
\left(1-\frac{T^2}{T_c^{\phi2}}\right)^2 \theta(T_c^\phi-T) \nn
& + & \fr{\pi^2T^4}{24}+\frac{\mu^2T^2}{48}
-\fr12\mu^2\chi_1^2
+\frac{1}{32\pi^2} \tilde m^4 \ln\frac{m_\chi^2}{\muT^2},
\ea
from which it follows that
\ba
f(T,q) & \stackrel{q\to\infty}{\approx} & f(T,0) + \frac{\tilde
m^4}{f^2}
\left(1-\frac{T^2}{T_c^{\phi2}}\right)^2 \theta(T_c^\phi-T) \nn
& + & \frac{\pi^2T^4}{24}+\frac{1}{16\pi^2}\tilde m^4
\ln\left(
\frac{fe^{\gamma+1/2}}{\sqrt{2}}\frac{q}{\tilde m^2 T}
\right). \la{largeq}
\ea
Hence $\partial_q^2 f(T,q) < 0$. We conclude that a (meta)stable
Q-matter state does not exist in the present system.

Consider then the plasma phase. That the plasma phase at $q<q_c$
satisfies the stability conditions in eqs.~\nr{stab1}, \nr{stab2}, only
proves that such a state is metastable against small perturbations. The
global minimum may still be elsewhere. Consider, in particular, an
arrangement where all the charge is moved into a subvolume $V' = rV$,
with $r<1$ (this of course means that there is no chemical equilibrium,
but this is not essential for the argument). Let $\delta
f(T,q)=f(T,q)-f(T,0)$. Then the original free energy density excess
with respect to an empty space is $\delta f(T,q)$, while the final is
$\delta f(T,q/r) r$. Now, for any initial $q$, $\delta f(T,q/r) r \to
0$ as $r\to 0$, since $\delta f(T,q/r)$ grows only logarithmically at
small $r$, see eq.~\nr{largeq}. Hence, even if $f(q/r)r$ grows at
$r\lsim 1$ and the original state is thus metastable, $f(q/r)r$ finally
starts to decrease for small enough $r$ and it is in principle
favourable to clump all the charge together in one place. Of course,
surface effects start to be important at very small $r$, so to be more
precise, one has to inspect the possible final configurations in more
detail. We thus turn to Q-balls.

\subsection{The stability of Q-balls \la{toystab}}

We start by considering the limit of large temperatures, $T\gg\tilde
m$, and small chemical potentials, $\mu \ll T$. Moreover, let us
subtract the contribution of the homogeneous symmetric plasma phase,
where $\chi_1=0$. Then it follows from eq.~\nr{4d1loop} that
\be
\bar V_\mu(\chi_1)\equiv
V_\mu(\chi_1)-V_\mu(0)\approx
-\fr12 \mu^2\chi_1^2 + \frac{T^4}{2\pi^2}
\biggl[
4 J_0(M_\chi/T)-\fr14 J_0(2M_\chi /T)
\biggr],
\ee
where $M_\chi=f\chi_1/\sqrt{2}$ and $J_0$ is from eq.~\nr{J0}.
Neglecting the change in the derivative term, the extra free energy of
any configuration $\chi_1({\bf r})$ is then given by
\be
F_\mu =
\int\! d^3 r \left[
\fr12 ( \partial_i\chi_1)^2+ \bar V(\chi_1) \right]. \la{extrF}
\ee

The properties of the saddle points of eq.~\nr{extrF} are discussed in
Appendix B. It follows from eqs.~\nr{qbp1}, \nr{qbp2} that in the
present model ($a_V=(\pi^2/24)T^4$), the Q-ball free energy, charge and
radius at $T\gg \tilde m$ are given by
\be
F_Q \approx 4.74 T Q^{3/4}, \quad \la{FQ}
Q  \approx 160 \left(\frac{\mu}{T}\right)^{-4}, \quad
R \approx \frac{\pi}{T} \left(\frac{\mu}{T}\right)^{-1}.
\ee

We now consider the stability of such Q-balls. The other metastable
alternative is the homogeneous plasma phase, so we have to compare with
$f(T,q)$ in eq.~\nr{plprops}. Let us write $F_Q = a T Q^{3/4}$, $Q= b
(\mu/T)^{-4}$.

Consider first a situation of thermal but not chemical equilibrium, in
which all the charge is in the Q-ball. Then, the Q-ball does not decay
into a homogeneous plasma phase, if $F_{\rm plasma}=V f_{\rm
plasma}(T,q)> F_Q$, or
\be
\fr65 a^{-1} Q^{5/4} > VT^3. \la{ball_eq}
\ee
This is the analogue of eq.~\nr{volume}, and tells that in a fixed
volume, Q-balls will evaporate at large enough temperatures.

On the other hand, note that what appears on the right-hand side of
eq.~\nr{ball_eq} is a comoving volume in units of temperature, which is
a constant in the cosmological context. Thus, in this case, if Q-balls
are thermally stable at some high temperature, they remain stable at
least as long as $T\gg\tilde m$. The thermal stability of a
distribution of Q-balls can be seen by replacing $V$ by their inverse
number density $n_Q^{-1}$.

Consider then a thermodynamical equilibrium situation where there is
chemical equilibrium, as well. Some of the charge will now be outside
the Q-ball, since there is a non-vanishing chemical potential. The
stability condition becomes
\be
\fr65\frac{Q^2}{VT^3} > \fr65\frac{(Q-Q')^2}{VT^3} + a (Q')^{3/4},
\ee
where the Q-ball charge $Q'$ is to be solved from an equality
following from chemical equilibrium:
\be
\frac{\mu}{T} \stackrel{\rm \mbox{\tiny plasma}}{=}
\frac{12}{5}\frac{Q-Q'}{VT^3} \stackrel{\rm \mbox{\tiny Q-ball}}{=}
\left(\frac{Q'}{b}\right)^{-1/4}. \la{mucondition}
\ee
However, it can easily be seen that when eq.~\nr{ball_eq} is strongly
enough satisfied, then the fact that there is chemical equilibrium does
not change the result. Indeed, it can be seen from eq.~\nr{mucondition}
that most of the charge resides in Q-balls in this limit,
\be
\frac{Q'}{Q} \approx 1 -
\frac{b^{1/4}}{2a} \left[\frac{VT^3}{(6/5)a^{-1} Q^{5/4}}\right]
\approx 1,
\ee
and Q-balls carry most of the free energy associated with the non-zero
net charge,
\be
\frac{F_{\rm plasma}}{F_Q}\approx
\frac{b^{1/2}}{4a^2}
\left[\frac{VT^3}{(6/5)a^{-1} Q^{5/4}}\right] \ll 1.
\ee
Moreover, provided that the Q-ball charge is large, Q-balls are small
in the sense that most of the space is in the plasma phase:
\be
\frac{V_Q}{V_{\rm plasma}} =
\frac{(4/3)\pi(R_QT)^3}{VT^3} \approx
\frac{10\pi^4 a}{9b^{3/4}}
\left[\frac{(6/5)a^{-1} Q^{5/4}}{VT^3}\right] \frac{1}{Q^{1/2}}
\ll 1,
\ee
if
\be
Q \gg 130
\left[\frac{(6/5)a^{-1} Q^{5/4}}{VT^3}\right]^2.
\ee

Even though chemical equilibrium does not essentially change the
constraint in eq.~\nr{ball_eq}, there is an important implication
following from eq.~\nr{mucondition}. Noting that the entropy
density is $s=dp/dT=\pi^2T^3/3$, one gets a simple relation between
the charge density in the plasma, and the charge
residing in a Q-ball:
\be
\left(\frac{q}{s}\right)_{\rm plasma} =
\frac{5}{4 \pi^2}\frac{\mu}{T} =
\frac{5 b^{1/4}}{4 \pi^2} (Q')^{-1/4} \approx 0.45 (Q')^{-1/4}.
\la{nBrelation}
\ee
In the cosmological context, this corresponds to a relation between the
baryon asymmetry in the plasma, and the baryon number residing in
Q-balls.

Finally, consider what happens at lower temperatures. The Q-ball free
energy will decrease for a while as $\sim T$, until at $T\lsim \tilde
m/f$, the symmetry breaking starts to play a role: eventually $T$ gets
replaced by $\tilde m/f^{1/2}$ in eq.~\nr{FQ}. At the same time, the
Q-part of the free energy of the plasma phase also decreases, initially
as $F_{\rm plasma} \sim (6/5)(Q^2/VT^3) T\sim T$. This behaviour is
valid up to $T\sim \tilde m/\pi$; at smaller temperatures one gets the
standard non-relativistic ideal gas behaviour $F_{\rm plasma} \sim QT
[\ln Q/(VT^3) +(3/2)\ln(T/\tilde m)]\sim T$. Hence the plasma phase
free energy decreases faster than $F_Q$ and a thermalized Q-ball in a
comoving volume becomes more unstable at smaller temperatures.

The considerations above concern an equilibrium situation. There are
also other relevant questions, such as, assuming that one is trapped in
a metastable homogeneous plasma phase and Q-balls would be stable, what
is the transition rate? Some progress in addressing this question has
been reported in~\cite{klee}, but we will not consider it here. In the
next Section we consider the opposite limit: taking a Q-ball which is
unstable, at which rate does it decay into the plasma phase?

\section{Baryon to cold dark matter ratio \la{sec:rate}}

Consider Q-balls in the cosmological context. In \cite{ks} it was shown
that Q-balls with large charges can be produced in the Early Universe.
The general idea is that in the AD scenario for baryogenesis, an
uniform condensate of the scalar field, carrying baryon number, is
formed as a result of inflation, CP-violation and baryon number
non-conservation \cite{ad}. This AD condensate is nothing but Q-matter,
described by an uniform scalar field with a time dependent phase. As we
have seen, Q-matter is unstable against Q-ball formation; an analysis
of the typical instability scales shows that charges of the order of
magnitude $Q > 10^{20}$ can be easily formed
(and do not violate experimental constraints~\cite{exp}).
In \cite{ks} it was
suggested that these stable Q-balls can play the role of the cold dark
matter. The ordinary, baryonic matter appears then as a result of
Q-ball evaporation. If true, baryonic and cold dark matter in the
Universe come from the same source. Moreover, dark matter is in fact
baryonic in this case, but baryons exist in the form of SUSY Q-balls
which do not participate in nucleosynthesis. In this scenario the ratio
of baryonic to cold dark matter can be computed and the aim of this
Section is to make such an estimate.

We assume that after the instability has taken place, the Universe is
filled
with baryonic Q-balls with charge\footnote{As is shown in \cite{ks},
the instability of an AD condensate occurs in a narrow range of scales, and
thus the Q-ball charge distribution is likely to be narrow, as well.}
$Q$ and number density $n_Q$, surrounded by a hot plasma at
temperature~$T$. We will also assume that only a negligible part of the
baryonic charge initially exists in the plasma. Some arguments if
favour of this assumption were put forward in \cite{emc2}, but its
complete check would require the solution of the non-linear problem of
Q-ball formation. In the following, we completely neglect the process
of merging of two Q-balls, since their concentration is so small that
the probability of their collision is absolutely negligible, as can be
checked a posteriori.

As a starting point, let us consider what the thermodynamical
equilibrium state corresponding to these initial conditions would be.
Then we will argue that in fact only thermal equilibrium can be
established in practice, while chemical equilibrium with respect to
baryon charge is not reached with these initial conditions.

If there is full thermodynamical equilibrium, then it follows from
eq.~\nr{mucondition} that already at high temperatures, the baryon
to photon ratio in the plasma phase would be
\be
\eta = \frac{n_B}{n_\gamma}\sim Q^{-1/4}. \la{nBs}
\ee
In full thermal equilibrium Q-balls become more unstable
at lower temperatures, see Sec.~\ref{toystab}, so that
with the experimental number $\eta \sim 10^{-10}$,
eq.~\nr{nBs} would imply the lower bound
$Q\gsim 10^{40}$. On the other hand,
if Q-balls account for a fraction $\Omega_{\rm DM}$ of the dark matter,
then their present number density is
$n_Q^{\rm (0)}=\rho_Q^{\rm (0)}/M_Q$, where
\be
\rho_Q^{\rm (0)} \approx 10^{-5} \Omega_{\rm DM} \,  m_p\, {\rm
cm}^{-3},
\ee
and $m_p$ is the proton mass. It follows that
\be
n_Q^{-1} T^3 \approx 3\times 10^8 g_*^{-1}
\frac{M}{m_p} \frac{Q^{3/4}}{\Omega_{\rm DM}}, \la{DM}
\ee
where $g_*$ is the number of massless degrees of freedom. If, for
instance, $g_*\sim 200, \Omega_{\rm DM}\sim 0.4, Q\sim 10^{40}$ and
$M\sim (1\ldots 10)$ TeV, then\footnote{To illustrate this number
density in cosmological units: $n_Q^{-1/3}\sim 10^{-2} R_H^{\rm ew}$,
where $R_H^{\rm ew}$ is the horizon radius at the electroweak scale
$T\sim 100$ GeV.} $n_Q^{-1} T^3 \sim 10^{39}\ldots 10^{40}$. Comparing
now with the LHS of eq.~\nr{ball_eq}, it is seen that such a density of
Q-balls would indeed be thermodynamically stable against evaporation
into the plasma phase. However, it is not clear how Q-balls with
charges as large as $10^{40}$ could be produced, and whether they would
indeed reach thermodynamical equilibrium in the cosmological
environment.

Now we are going to take into account the expansion of the Universe.
First of all, the intrinsic temperature of Q-balls may be
different from the temperature of the surrounding plasma,
if the system is not fully thermalized.
Then Q-balls are
heated up because of collisions with particles from the plasma. The
energy transfer to the scalar particles in the condensate
from which the Q-ball is
built, has been estimated in \cite{ks} (we assume that all coupling
constants are of order unity),
\be
\frac{d F}{d t} \sim R_Q^2 T^3 \sigma n_\phi \delta R \delta E 
\sim T^2 (\frac{T}{M})^3 Q^{1/4}. \la{Trate}
\ee
Here $R_Q^2 T^3$ counts the initial flux of particles in the
plasma on the Q-ball, $\delta E \sim \omega$ gives the energy transfer to
a scalar particle in the condensate, $\delta R \sim T/M^2$ gives the
thickness of the Q-ball layer 
in which plasma particles can penetrate (due to $m_\phi\sim\phi\lsim T$),
$n_\phi \sim \phi^2 \omega \sim T^2 M Q^{-1/4}$ is the 
charged scalar particle number
density in this layer [eq.~\nr{omega}], 
$\sigma \sim 1/(\omega T)$ is the typical
cross-section\footnote{In
\cite{ks} this cross-section was incorrectly taken to be $\sigma \sim
1/T^2$. Also, the estimate of the transmitted energy $\delta E$
contains a misprint: 
the power $2$ must be omitted from the corresponding
expression.} of the interaction of an energetic particle from the plasma
with a condensate scalar with a small energy $\sim \omega$, 
and we assumed $T\gg M$. 
The estimate in eq.~\nr{Trate} contains only the effect
of particles interacting strongly (at tree-level) with 
the condensate, so that it should represent a lower bound
of the thermalization rate.

The energy transfer is most efficient at high temperatures.
Requiring that the change
of Q-ball free energy is of the order of its equilibrium value $F\sim T
Q^{3/4}$, we find that a zero-temperature Q-ball can be thermalized
if the initial temperature of the plasma satisfies the requirement
\be
T_{\rm in} > M \left(\frac{M}{M_0}\right)^{1/2}Q^{1/4},
\ee
where $M_0=(0.30/\sqrt{g_*}) m_{\rm Pl}\sim 3\times 10^{17}$ GeV. For
typical values of $M \sim 1-10$ TeV and $Q\sim 10^{20}-10^{30}$,
$T_{\rm in} \gsim 1\ldots 10^{4}$ GeV, which is true
for most inflationary models.

Next, we would like to determine the amount of baryons which
evaporated from thermalized Q-balls. From the discussion in
Sec.~\ref{sec:qballs},
we know that we can associate with Q-balls a chemical potential $\mu
\sim M(T) Q^{-1/4}$. In the case that the surrounding plasma has the
same chemical potential, Q-balls are in chemical equilibrium and their
charge does not change. In other words, the rate of their evaporation
coincides with the rate of charge accretion on the Q-ball. The upper
limit on the accretion can be easily computed: the rate of accretion
cannot be larger than the total baryonic flux through the Q-ball
surface. So,
\be
\frac{dQ}{dt} = - D(\mu_Q-\mu_{\rm plasma})T^2 4 \pi R_Q^2
\stackrel{\mu_{\rm plasma}\ll \mu_Q}{\sim} - D\frac{T^2 Q^{1/4}}{M(T)},
\label{evap}
\ee
where the coefficient $D\lsim 1$ shows the deviation of the real
evaporation rate from the upper bound. At $T>m_\phi$ ($m_\phi$ is the
mass of squarks, suppressed by some coupling constants with respect to $M$),
the evaporation rate must be parametrically the same
as the upper bound, since the scalar particles building up the Q-ball
appear with a large number density $\sim T^3$ also in the plasma
phase\footnote{In case some of the Lagrangian squark mass parameters
are small or negative, this regime extends down to the critical temperature
of the electroweak phase transition.}.
Hence, the rate of Q-ball evaporation in
a plasma with a vanishing chemical potential ($\mu_{\rm plasma}=0$) at
$T>m_\phi$ is given by eq.~(\ref{evap}) with $D\sim 1$. At $T<m_\phi$,
the charge in the plasma sits mainly in light fermions; the probability
of squark emission from a Q-ball is suppressed by the Boltzmann exponent
$\exp(-m_\phi/T)$. Therefore, the emission of heavy squarks from the
Q-ball surface can be neglected. The main process which is allowed is
squark-squark transition into two quarks, $\phi \phi \rightarrow qq$.
It may occur through gluino exchange with a cross-section of the order
of $\sigma \sim \beta/m_\phi^2$ (see also a similar discussion in
\cite{kstt}), where $\beta$ is some combination of the coupling constants.
Thus, we expect $D \sim \beta T^2/m_\phi^2$ at $T<m_\phi$.

By integrating eq. (\ref{evap}) (we assume a radiation dominated
Universe) we find that Q-ball evaporation is negligible at $T<m_\phi$
($D$ is essentially replaced by $\beta$ in eq.~\nr{qfinal}).
The charge $Q_T$ of a Q-ball at $T\sim m_\phi$ is related to its
initial charge $Q_0$ as
\be
Q_T \sim Q_0\left( 1- D \frac{M_0}{M Q_0^{3/4}}
\ln(\frac{M}{m_\phi})\right)^{4/3}.
\label{qfinal}
\ee
Most of the charge is evaporated at low temperatures near
$T\sim m_\phi$.

{}From eq. (\ref{qfinal}) we conclude that Q-balls survive evaporation
if their initial charge is
\be
Q_0 \gsim \left(D \frac{M_0}{M} \ln(\frac{M}{m_\phi})\right)^{4/3}.
\label{stab}
\ee
If, for example, $M\sim 1$ TeV, then all Q-balls with $Q_0 > 10^{20}$
stay till the present time. It is interesting to note that these values
of $Q$ are naturally produced from the decay of an AD condensate.

Now, if eq.~(\ref{stab}) is satisfied, then the amount of charge
evaporated
from an individual Q-ball is of the order of
\be
\Delta Q \sim D \frac{M_0}{M} \ln(\frac{M}{m_\phi})  Q_0^{1/4}.
\ee
Thus, the number density of baryons $n_B$
in the plasma is related to the number
density of Q-balls $n_Q$ as
\be
n_B \sim n_Q \Delta Q.
\ee
Using then $n_Q$ from eq.~\nr{DM} together with the
entropy density  $s \approx (2/45) g_* \pi^2 T^3$,
and assuming that Q-balls give all the cold dark matter in the
Universe with $\Omega_{\rm DM} \sim 0.4$,
the baryon to photon ratio is computable:
\be
\eta = \frac{n_B}{n_\gamma} \sim 10^{-8} \frac {M_0 m_p}{M^2
Q_0^{1/2}} \ln(\frac{M}{m_\phi}).
\ee
The correct ratio $\eta\sim 10^{-10}$ appears quite naturally with the
choice of, e.g., $M\sim 1-10$ TeV and a corresponding initial charge
$Q_0 \sim 10^{28}-10^{22}$. These values are quite plausible from the
mechanism of Q-ball formation.

Since Q-balls interact very rarely, they should not take part in light
element nucleosynthesis. From the point of view of structure formation,
they are completely equivalent to weakly interacting cold dark matter.
They thus behave in a somewhat similar way as quark
nuggets~\cite{nucsynth}.

\section{Conclusions \la{sec:concl}}

Supersymmetric non-topological solitons in theories with flat
directions of the effective potential have a number of interesting
properties at high temperatures. Provided that their
charge is large enough, Q-balls can exist at any temperature.
Unlike for ``classic" Q-balls with $M_Q\sim Q^1$, 
the ground state of a system with a finite density of the
charge is not translationally invariant and contains Q-balls.
The state of Q-matter (Bose-Einstein condensate) appears to be always
unstable and it decays into Q-balls.

These properties have interesting implications for cosmology.
Depending on the mechanism of Q-ball formation, thermal and chemical
equilibrium may or may not be reached. If it is reached, an equilibrium
distribution of Q-balls with charges $Q\gsim 10^{40}$ could account
for both the net baryon asymmetry needed for
nucleosynthesis, and for the cold dark matter. In this picture,
the dark matter is in fact baryonic, with the baryon charge
distributed between ordinary baryons with $n_B/n_\gamma \sim 10^{-10}$,
and squarks packed inside dark matter Q-balls.

On the other hand, for charges $Q\sim 10^{22}\ldots 10^{28}$ which
are more likely from the point of view of Q-ball formation, only
thermal equilibrium is reached. But even then, it turns out that
the right orders of magnitude for the baryon asymmetry in ordinary
baryons and for the dark matter in Q-balls, can be reached provided
that supersymmetry is broken at a low energy scale
$M\sim 1\ldots 10$ TeV. Thus Q-balls represent an appealing candidate
for the cold dark matter, offering a new picture in which the
origin of ordinary baryons and the dark matter is the same.

\section*{Acknowledgements}

We are grateful to A. Kusenko for useful discussions.

\appendix
\appendix
\renewcommand{\thesection}{Appendix~~\Alph{section}:}
\renewcommand{\thesubsection}{\Alph{section}.\arabic{subsection}}
\renewcommand{\theequation}{\Alph{section}.\arabic{equation}}

\section{A model effective potential}

In this appendix, we derive the 1-loop effective potential
of the theory considered in Sec.~\ref{sec:model} for given
$\mu,T$, both at small and large values of the fields.
These two cases have to be treated separately, as the
theory requires a resummation for small values of the fields.

At 1-loop level, one has to consider renormalization.
To be specific, the running of the fields and parameters
is at 1-loop level determined by
\ba
\overline\mu\frac{d}{d\overline\mu} f^2 & = & \frac{5}{16\pi^2}f^4,
\quad\quad
\overline\mu\frac{d}{d\overline\mu} \tilde m^2 = \frac{3}{16\pi^2}
f^2 \tilde m^2, \\
\overline\mu\frac{d}{d\overline\mu} \phi^*\phi & = &
-\frac{2}{16\pi^2}f^2 \phi^*\phi, \quad\quad
\overline\mu\frac{d}{d\overline\mu} \chi^*\chi =
-\frac{1}{16\pi^2}f^2 \chi^*\chi.
\ea
Here $\phi^*\phi$, $\chi^*\chi$ denote the renormalized fields
squared appearing in the Lagrangian, and {\em not} any composite
operators. As a result, the parameters and
fields at scale $\overline\mu$ are
\ba
\phi^*\phi(\overline\mu)  & \approx &
\phi^*\phi(\overline\mu_0)\left(\frac{\overline\mu}{\overline\mu_0}
\right)^{-\frac{2f^2}{16\pi^2}},\quad
\chi^*\chi(\overline\mu) \approx
\chi^*\chi(\overline\mu_0)\left(\frac{\overline\mu}{\overline\mu_0}
\right)^{-\frac{f^2}{16\pi^2}}, \\
\tilde m^2(\overline\mu)  & \approx  &
\tilde m^2(\overline\mu_0)\left(\frac{\overline\mu}{\overline\mu_0}
\right)^{\frac{3f^2}{16\pi^2}}, \quad
f^2(\overline\mu)
\approx \frac{16\pi^2}{5\ln(\Lambda/\overline\mu)},
\ea
where $\Lambda=\overline\mu_0\exp[16\pi^2/5f^2(\overline\mu_0)]$.

The theory can then be defined in terms of two parameters, the values
of $\tilde m^2(\overline\mu_0)$, $f(\overline\mu_0)$,
at some value $\overline\mu_0$
of the $\msbar$ scale parameter. We choose
$\overline\mu_0\equiv \tilde m_0$, where
$\tilde m_0 = \tilde m (\tilde m_0)$. Then all the
dimensionful quantities, such as the temperature $T$,
are measured in units of $\tilde m_0$, and the results
only depend on $f(\tilde m_0)$. To have a weakly
coupled theory up to very high scales, we arbitrarily
choose $f(\tilde m_0)=0.5$.

The most convenient way to
implement the resummation needed at small
values of the fields, is to construct
an effective 3d theory describing the
thermodynamics~\cite{generic}--\cite{erice}.
On the other hand, for large values of $\chi$, no resummation
is needed and the effective 3d theory is not valid any more.
Then one can use the standard 4d finite temperature
1-loop potential. Let us discuss the 3d theory first.

\subsection{The 3d effective theory and $V(\phi,\chi)$ at
small $\phi,\chi$\la{sec:3d}}

Let us denote
\be
M_\phi = \frac{1}{\sqrt{2}} f\phi_1,\quad
M_\chi = \frac{1}{\sqrt{2}} f\chi_1.
\ee
The effective theory constructed is
valid for $M_{\phi},M_{\chi}\ll \pi T$,
and implements the resummation needed in the regime
$M_{\phi,\chi}\lsim fT$.
%The 3d theory would in principle also allow non-perturbative
%lattice simulations of the thermodynamical properties of the theory.

The derivation of the effective theory proceeds by
computing the $n$-point functions in the original theory
and in the effective theory, and by matching the parameters
of the effective theory such that the original results are
reproduced. In the present context, one has to consider the
2-point functions of the bosonic fields at non-zero momenta
to get the field normalizations, and the 2- and 4-point
functions at zero momenta. The field normalizations arise
from fermionic loops alone. The zero-momentum
correlators arise both from bosonic and fermionic
loops, and the most convenient way of deriving them
is the effective potential, expanded in terms of
masses to the required order. We will also expand
in $\mu$, assuming $\mu \ll \pi T$, although this would
not be necessary. We then assume that parametrically
$f\phi,f\chi,\mu\sim fT$
and work at 1-loop level to order $f^4$. Thus we need
the first five terms (up to $B^4$) from the expansion
\be
\fr12 \ln \det (A+B) =
\fr12 \ln \det A+
\sum_{n=1}^\infty \frac{(-1)^{n+1}}{2 n}\tr (A^{-1}B)^n.
\la{logexpansion}
\ee
In the bosonic case, the matrices are
\ba
A^{-1} & = &  \frac{1}{p^2} \mathop{\rm diag} (1,1,1,1), \nn
B & = & \left(
\begin{array}{cccc}
\tilde m^2+\fr12 M_\phi^2 + M_\chi^2 & 0 & 0 &  0\\
0 & -\tilde m^2 + \fr32 M_\phi^2+M_\chi^2 & 2 M_\phi M_\chi & 0 \\
0 & 2 M_\phi M_\chi & -\mu^2 + M_\phi^2 & -2\mu p_0 \\
0 & 0 & 2 \mu p_0 & -\mu^2 + M_\phi^2
\end{array}
\right), \hspace*{1cm}
\ea
where $p^2=p_0^2+\bfp^2$ and $p_0$ is a bosonic Matsubara frequency.
In the fermionic case,
\be
A^{-1} = \frac{1}{p^2}\left(
\begin{array}{cc}
-i\slash\!\!\! p & 0 \\
0 & -i\slash\!\!\! p
\end{array}
\right), \quad
B = \left(
\begin{array}{cc}
-\frac{\mu}{2}\gamma_0\gamma_5 + M_\chi & M_\phi \\
M_\phi & \frac{\mu}{2}\gamma_0\gamma_5
\end{array}
\right).
\ee
The fermionic matrices have been written in the basis of the Majorana
fermions
\be
\tilde\psi_\phi=\left(
\begin{array}{c}
\psi_\phi\\
\overline\psi_\phi
\end{array}
\right),\quad
\tilde\psi_\chi=\left(
\begin{array}{c}
\psi_\chi\\
\overline\psi_\chi
\end{array}
\right), \la{majorana}
\ee
where we used the standard notation of Ref.~\cite{hk}.
There is an overall minus sign
in eq.~\nr{logexpansion} for fermions,
and only even powers of $B$
contribute due to the momentum integration.

The basic integrals appearing in the evaluation of the traces are
\ba
& & \Tint{p_b}' \frac{1}{p^2} = \frac{T^2}{12},
\quad
\Tint{p_b}' \frac{1}{(p^2)^2} = \frac{1}{16\pi^2}
\left[\frac{1}{\epsilon}+L_b(\overline\mu)\right], \quad
\Tint{p_b}' \frac{1}{(p^2)^3} = \frac{\zeta(3)}{128\pi^4T^2}, \\
& & \Tint{p_f} \frac{1}{p^2} = -\frac{T^2}{24},
\quad
\Tint{p_f} \frac{1}{(p^2)^2} = \frac{1}{16\pi^2}
\left[\frac{1}{\epsilon}+L_f(\overline\mu)\right], \quad
\Tint{p_f} \frac{1}{(p^2)^3} = \frac{7\zeta(3)}{128\pi^4T^2}, \hspace*{0.5cm}
\la{ints}
\ea
where $p_b,p_f$ are the bosonic and fermionic Matsubara momenta,
a prime means that the zero Matsubara mode is omitted, and
\be
L_b(\overline\mu)= 2 \ln \frac{\overline\mu}{\muT}, \quad
L_f(\overline\mu)= 2 \ln \frac{4\overline\mu}{\muT}, \quad
\overline\mu_T = 4\pi e^{-\gamma} T.
\ee
The other integrals appearing, of the
form $\int\!\!\!\!\!\raise-0.2ex\hbox{$\Sigma$}\, (p_0^2)^n/(p^2)^m$,
can be derived
from those in eq.~\nr{ints} by taking derivatives
with respect to $T$.

As a result, the 3d finite temperature
effective theory describing the thermodynamics of the original 4d theory
is defined by the action
\ba
S_{\rm 3d} & = &
\frac{V}{T}
\left[-\frac{\pi^2}{12}T^4-\frac{5}{24}\mu^2T^2+
\frac{7}{192\pi^2}\mu^4+
\frac{\tilde m^4(\overline\mu_T)}{f^2(\overline\mu_T)} \right]
\nonumber \\
& + & \int\! d^3x\, \biggl[
\fr12 (\nabla\phi_1)^2+\fr12(\nabla\phi_2)^2+
(\nabla\chi)^*(\nabla\chi) \nonumber \\
& & +\fr12 m_1^2 \phi_1^2+\fr12 m_2^2\phi_2^2 +
m_\chi^2 \chi^*\chi \nonumber \\
& & +
\fr14 \lambda_\phi(\phi_1^2+\phi_2^2)^2+
\fr12 \lambda_m (\phi_1^2+\phi_2^2)\chi^*\chi+
\lambda_\chi(\chi^*\chi)^2 \biggr]. \la{3daction}
\ea
Here the fields are related to the 4d fields by
\be
(\phi\phi)_{\rm 3d} = \frac{1}{T}
(\phi\phi)_{\rm 4d}\Bigl(\fr14 \overline\mu_T\Bigr),
\quad
(\chi\chi)_{\rm 3d} = \frac{1}{T}
(\chi\chi)_{\rm 4d}\Bigl(\fr14 \overline\mu_T\Bigr).
\ee
The parameters are given by
\ba
m_1^2 & = &
-\tilde m^2\left(2^{-4/3}\overline\mu_T\right)
-\frac{f^2\mu^2}{16\pi^2}+
\frac{f^2T^2}{4}, \nn
m_2^2 & = &
\tilde m^2\left(2^{-4/3}\overline\mu_T\right)
-\frac{f^2\mu^2}{16\pi^2}+
\frac{f^2T^2}{4}, \nn
m_\chi^2 & = &
-\mu^2\left(1-\frac{f^2}{16\pi^2}\right)+\frac{f^2T^2}{8}, \nn
\lambda_\phi & =  & \fr14 T f^2\left(2^{8/5} \muT\right), \quad
\lambda_m = T f^2\left(2^{2/5}\muT\right), \quad
\lambda_\chi = T  \frac{f^4}{8\pi^2}\ln2. \la{3dprms}
\ea
Note, in particular, that around the two ``symmetry restoring''
temperatures [$T_c^{\phi} = 2\tilde m/f$ where
$m_1^2\sim 0$, and $T_c^{\chi} = 2\sqrt{2} \mu/f$ where
$m_\chi^2 \sim 0$],
one has $\pi T\gg \{\tilde m, \mu\}$ for a small coupling $f$,
so that the construction of the effective theory is well
convergent.

To go further with resummed perturbation theory, one can compute
the effective potential
in the theory defined by the action in eq.~\nr{3daction}.
The general pattern can be seen already from the tree-level potential,
\be
V_{\rm tree} = {\rm const.} +
\fr12 m_1^2\phi_1^2 +
\fr12 m_\chi^2 \chi_1^2 +
\fr14 \lambda_\phi \phi_1^4+
\fr14 \lambda_m \phi_1^2\chi_1^2+
\fr14 \lambda_\chi \chi_1^4. \la{3dtree}
\ee
This tree-level potential of the 3d theory corresponds to
the dominant thermal screening contributions
in the 4d 1-loop potential.
The 3d 1-loop correction is
\be
V_{\rm 1-loop} =
-\frac{1}{12\pi}\sum_{i=\phi_2,\chi_2,\pm} m_i^3, \la{3d1loop}
\ee
where
\ba
m_{\phi_2}^2 & = &
m_2^2+\lambda_\phi \phi_1^2+\fr12 \lambda_m \chi_1^2, \quad
m_{\chi_2}^2 =
m_\chi^2+\lambda_\chi \chi_1^2+\fr12 \lambda_m \phi_1^2, \nn
m_{\pm}^2 & = & \fr12\left[m_{\phi_1}^2+m_{\chi_1}^2
\pm\sqrt{(m_{\phi_1}^2-m_{\chi_1}^2)^2+4 m_{\phi\chi}^2}
\right], \nn
m_{\phi_1}^2 & = &
m_1^2+3\lambda_\phi \phi_1^2+\fr12 \lambda_m \chi_1^2, \quad
m_{\chi_1}^2 =
m_\chi^2+3\lambda_\chi \chi_1^2+\fr12 \lambda_m \phi_1^2, \nn
m_{\phi\chi}^2 & = & \lambda_m \phi_1\chi_1.
\ea
The 2-loop corrections can also be readily computed, but they do
not affect our conclusions.

\subsection{The effective potential $V(\phi,\chi)$
at large $\chi$\la{sec:4d}}

At large field values, $M_\chi \gsim \pi T$, the 3d effective
theory breaks down. Then one needs
the full finite temperature 1-loop effective potential,
but on the other hand resummation is not needed.
Moreover, the $\phi_1$-symmetry is restored, $M_\phi=0$.
We assume also that the chemical potential is small, $\mu\ll M_\chi$,
which is well justified in the region considered.

The tree-level potential for $M_\phi=0$ is
\be
V_{\rm tree} = \frac{\tilde m^4(\overline\mu)}{f^2(\overline\mu)}
-\fr12 \mu^2 \chi_1^2. \la{4dtree}
\ee
The mass spectrum entering the 1-loop potential
consists of
two massless scalar degrees of freedom $\chi_1,\chi_2$
and one massless Majorana spinor $\tilde \psi_\chi$,
together with the massive
scalar particles $\phi_1,\phi_2$ and the Majorana
fermion $\tilde\psi_\phi$. The masses are
\be
m_{\psi_\phi} = M_\chi, \quad
m_{\phi_1}^2 = -\tilde m^2+ m_{\psi_\phi}^2, \quad
m_{\phi_2}^2 = \tilde m^2+ m_{\psi_\phi}^2.
\ee
To order $(\mu/M_\chi)^2$,
the renormalized 1-loop contribution is then
\ba
V_{\rm 1-loop} & = &
\left[
-2\left(1+\fr78\right)\frac{\pi^2T^4}{90}-\frac{9}{48}\mu^2T^2
\right]
%+\frac{5}{128\pi^2}\mu^4
\nn
& + &
\sum_{i=1,2}\left[
%% FOLLOWING LINE CANNOT BE BROKEN BEFORE 72 CHAR
%% FOLLOWING LINE CANNOT BE BROKEN BEFORE 72 CHAR
-\frac{m_{\phi_i}^4}{64\pi^2}\left(\ln\frac{\overline\mu^2}{m_{\phi_i}^2 
 }+
\fr32\right)+\frac{T^4}{2\pi^2} J_0(m_{\phi_i}/T)
\right] \nn
& + & (-2)
\left[
-\frac{m_{\psi_\phi}^4}{64\pi^2}\left(\ln
\frac{\overline\mu^2}{m_{\psi_\phi}^2}+
\fr32\right)+\frac{T^4}{2\pi^2} \left(
\fr18 J_0(2 m_{\psi_\phi}/T)-J_0(m_{\psi_\phi}/T)\right)
\right] \nn
& + & \frac{\mu^2}{2}
\left[
-\frac{m_{\psi_\phi}^2}{16\pi^2}\ln
\frac{\overline\mu^2}{m_{\psi_\phi}^2}+
\frac{T^2}{2\pi^2} \left(
\fr12 I_0(2 m_{\psi_\phi}/T)-I_0(m_{\psi_\phi}/T)\right)
\right]. \la{4d1loop}
\ea
The integrals appearing here are
\ba
J_0(y) & = &
\int_0^\infty \! dx\, x^2 \ln \left(
1-e^{-\sqrt{x^2+y^2}}
\right), \la{J0} \\
I_0(y) & = &
\int_0^\infty \! dx
\frac{x^2}{\sqrt{x^2+y^2}}\frac{1}{e^{\sqrt{x^2+y^2}}-1}.
\ea
In the regime $M_\chi\lsim T$,
the convergence of the 1-loop potential
can be improved by a resummation:
\be
V'_{\rm 1-loop} =
V_{\rm 1-loop} + \sum_{i=1,2} \frac{T}{12\pi}
\left( m_{\phi_i}^3-\overline m_{\phi_i}^3\right),
\ee
where $\overline m_{\phi_i}^2=m_{\phi_i}^2+ f^2T^2/4$.

\section{Q-ball properties}
\label{sec:qball}

In this appendix, we briefly review the properties of Q-balls
in theories with flat directions. The high temperature limit
of Q-balls in the theory considered in Sec.~\ref{sec:model}
follows as a special case.

Consider the action for the length $\chi_1$ of
the complex U(1) scalar field $\chi$ (the extremum
solutions considered do not depend on the phase of $\chi$),
\be
F_\mu =
\int\! d^3 r \left[
\fr12 ( \partial_i\chi_1)^2-\fr12\mu^2\chi_1^2+
V(\chi_1) \right]. \la{Fmuac}
\ee
Here it has been assumed that the derivative part of the
effective action remains the same as at tree-level
and only the potential changes.
The potential is assumed to be normalized such that $V(0)=0$.
Let us scale eq.~\nr{Fmuac}
into a dimensionless form by $V(\chi_1)=a_V g(\tilde \chi)$,
$\chi_1=a_\chi\tilde \chi, \mu=a_\mu\tilde\mu, r=a_R\tilde r, 
F=a_F\tilde F, Q=a_Q\tilde Q$, with
$a_\mu=\sqrt{a_V}/a_\chi,a_R=a_\chi/\sqrt{a_V},
a_F=a_\chi^3/\sqrt{a_V},a_Q=a_\chi^4/a_V$.
The scaling of $V(\chi_1)$ is chosen so that
$g(0)=0,g(\infty)=1$. It then follows that the
spherically symmetric extrema of the action
in eq.~\nr{Fmuac}, required by eq.~\nr{effaction},
satisfy
\be
\frac{d^2\tilde\chi}{d\tilde r^2}+
\frac{2}{\tilde r}\frac{d\tilde\chi}{d\tilde r}=
-\tilde \mu^2\tilde\chi + g'(\tilde \chi);\quad
\tilde\chi'(0)=0;\quad
\tilde\chi(\infty)=0. \la{qballeq}
\ee
%The solutions of eq.~\nr{qballeq} are the precise
%analogues of the zero temperature Q-balls.

The main properties of the solution of eq.~\nr{qballeq}
for small $\tilde\mu$
are familiar from the zero temperature context,
and do not depend at all on the functional form of $g(\tilde\chi)$
at small $\tilde\chi$. To see this,
assume that the solution scales so that
$\tilde F_\mu = C \tilde\mu^{-\alpha}$.
Then
\be
\tilde Q
= \tilde\mu\int\!d^3\tilde r\tilde\chi^2 \
= -\frac{\partial\tilde F_\mu}{\partial\tilde\mu}
= C\alpha \tilde\mu^{-\alpha-1},
\la{tildeQ}
\ee
and
\be
\tilde F_Q = \tilde F_\mu +\tilde\mu\tilde Q =
(1+\alpha^{-1})(\alpha C)^{\frac{1}{1+\alpha}}
\tilde Q^{\frac{\alpha}{1+\alpha}}. \la{tildeFQ}
\ee
The parameters $\alpha, C$ can now be easily derived.
At large $\tilde\chi$
(small $\tilde r$), the potential $g(\tilde \chi)$ equals
unity, and the equations of motion can be solved analytically:
\be
\tilde\chi(\tilde r) = \tilde \chi_0
\frac{\sin\!\tilde\mu\tilde r}{\tilde \mu \tilde r}. \la{tchir}
\ee
The scale of $\tilde\chi_0$ is determined by when
$-(1/2)\tilde\mu^2\tilde\chi^2$ compensates for the unity in
$g(\tilde\chi)$.
It follows that $\tilde\chi_0\sim \tilde\mu^{-1}$. The
radius of the Q-ball is determined by $\tilde\chi(\tilde R)\sim 1$,
i.e.\ $\tilde R \sim \pi\tilde\mu^{-1}$. Thus
\be
\tilde Q = \tilde \mu \int\!d^3\tilde r\tilde\chi^2
\sim \tilde\mu\tilde R^3\tilde\chi_0^2 \sim \tilde \mu^{-4},
\la{Qint}
\ee
and it follows from eq.~\nr{tildeQ} that $\alpha=3$.
To determine the constant $C$, note that
$g(\kappa/\tilde\mu)\to \theta(\kappa)$
with $\theta(0)=0$, when $\tilde\mu\to 0$.
Eq.~\nr{qballeq} can then be integrated
from $\tilde R-\epsilon$ to $\tilde R+\epsilon$,
and one finds $\tilde\chi_0=\sqrt{2}\pi/\tilde\mu$.
One then gets from eqs.~\nr{tildeQ}, \nr{tchir}
that $C=4\pi^4/3$.

We can now go back to the original
dimensionful units, to find that
\be
R=(1/\sqrt{2}) a_V^{-1/4} Q^{1/4}, \quad
\chi_{1}^{\rm max} = a_V^{1/4}Q^{1/4}, \quad
R \mu = \pi,  \la{qbp1}
\ee
\be
F_Q = (4\pi\sqrt{2}/3)a_V^{1/4} Q^{3/4}, \quad
Q = 4\pi^4 a_V \mu^{-4}. \la{qbp2}
\ee
These are in accordance with eqs.~\nr{props1}, \nr{props2},
where $a_V=M^4$. In the model of Sec.~\ref{sec:model} at
very high temperatures, $a_V=(\pi^2/24)T^4$.

%%%%%%%%%%%%%%%%%%%%%%%%%%%%%%%%% FIGURE
\begin{figure}[t]

\vspace*{-1cm}

\centerline{ %\hspace{-3.3mm}
\epsfxsize=9cm\epsfbox{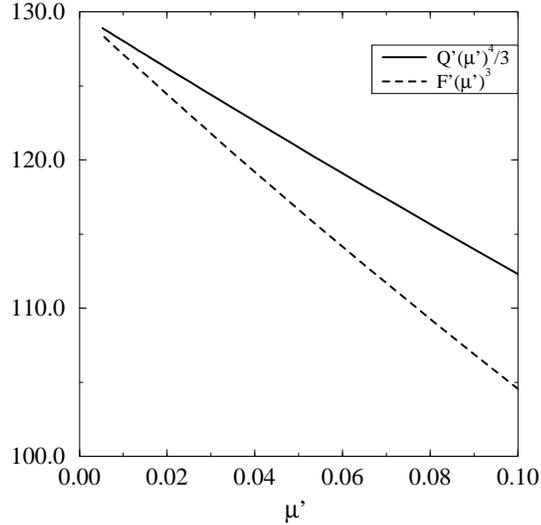}}

\vspace*{-4.5cm}

\caption[a]{The free energy $\tilde F_\mu$ and
the charge $\tilde Q$ as a function of the scaled
chemical potential $\tilde\mu$ (denoted by $F',Q',\mu'$
in this figure) for the Q-ball solution, in the limit of high
temperatures
and a small chemical potential.
The extrapolated constant at $\tilde\mu\to0$ is $C\approx129.87(1)$,
in accordance with the analytic result $C=4\pi^4/3$.
The slopes are such that $\tilde
F_\mu=C\tilde\mu^{-3}(1-2.1\tilde\mu)$,
$\tilde Q=3 C\tilde\mu^{-4}(1-1.4\tilde\mu)$.}
\la{fig:5}
\end{figure}
%%%%%%%%%%%%%%%%%%%%%%%%%%%%%%%%%%%%

Finally, note that in contrast to the leading terms,
the first corrections to the asymptotic behaviour
in eqs.~\nr{qbp1}, \nr{qbp2} depend on the detailed form
of the potential $g(\tilde\chi)$. As an example,
consider the model in Sec.~\ref{sec:model}.
With $a_V=(\pi^2/24)T^4$ and choosing
$\tilde\chi=M_\chi/T=f\chi_1/(\sqrt{2}T)\equiv a_\chi^{-1}\chi_1$,
we get
\be
g(\tilde\chi)=
1+\frac{48}{\pi^4}J_0(\tilde\chi)-\frac{3}{\pi^4}J_0(2\tilde\chi).
\ee
Here $J_0$ is from eq.~\nr{J0}. The
numerical values following from the
solution of eq.~\nr{qballeq}
with this $g(\tilde\chi)$ are shown in Fig.~\ref{fig:5}.
Since $\tilde\mu = (a_\chi/\sqrt{a_V})\mu = (4\sqrt{3}/f\pi)(\mu/T)$,
it is seen from Fig.~\ref{fig:5}
that the asymptotic
estimates in eqs.~\nr{qbp1}, \nr{qbp2} are
reliable within $\sim 10$\% at least up to
$\tilde \mu\lsim 0.05$, or $\mu/T\lsim 0.02 f$. This
corresponds to charges $Q\gsim 10^{9} f^{-4}$.

\end{document}